\documentclass{article}
\usepackage{amsmath,amsfonts,amssymb,amsthm}
\usepackage{latexsym}
\usepackage{epsfig,overpic}
\usepackage{stmaryrd}
\usepackage{color}
\usepackage{graphicx}
\usepackage{wrapfig}
\usepackage{xcolor}
\usepackage{hyperref}

\newtheorem{remark}{Remark}[section]

\providecommand{\keywords}[1]{\textbf{\textit{Keywords:}} #1}
\newcommand{\rev}[1]{\textcolor{black}{{#1}}}
\let\eps\varepsilon
\def\ds{\mathrm{d}\mathbf{s}}

\newcommand{\bM}{\mathbf M}

\newcommand{\bV}{\mathbf V}
\newcommand{\bU}{\mathbf U}
\newcommand{\bF}{\mathbf F}

\newcommand{\bg}{\mathbf g}
\newcommand{\bn}{\mathbf n}

\newcommand{\bv}{\mathbf v}
\newcommand{\bl}{\mathbf l}

\newcommand{\bw}{\mathbf w}
\newcommand{\bx}{\mathbf x}
\newcommand{\bq}{\mathbf q}
\newcommand{\bt}{\mathbf t}
\newcommand{\by}{\mathbf y}

\newcommand{\F}{\mathcal F}

\newcommand{\T}{\mathcal T}

\newcommand{\cL}{\mathcal L}
\newcommand{\cR}{\mathcal R}

\newcommand{\Div}{\mbox{\rm div}}

\definecolor{darkred}{rgb}{.7,0,0}

\definecolor{green}{rgb}{0,0.7,0}

\def\dO{{\partial\Omega} }
\def\dG{{\partial\Gamma} }

\def\nath{\nabla_{\Gamma_h}}

\begin{document}

\title{A hybrid finite volume -- finite element method for bulk--surface coupled problems}
\author{
Alexey Y. Chernyshenko\thanks{
Institute of Numerical Mathematics, Russian Academy of Sciences, Moscow 119333}
\and
Maxim A. Olshanskii\thanks{Department of Mathematics, University of Houston, Houston, Texas 77204-3008 {\tt molshan@math.uh.edu}}
\and
Yuri V. Vassilevski\thanks{
Institute of Numerical Mathematics, Russian Academy of Sciences, Moscow 119333 and Sechenov University, Moscow 119991}
}

\date{}
\maketitle

\markboth{}{}

\begin{abstract}
The paper develops a hybrid method for solving a system of  advection--diffusion equations
in a bulk domain coupled to advection--diffusion equations on an embedded surface. A monotone nonlinear finite volume method for equations posed in the bulk is combined with a trace finite element method for equations posed on the surface.
In our approach, the surface is not fitted by the mesh and is allowed to cut through the background  mesh in an arbitrary way.   Moreover,  a triangulation of the surface into regular shaped elements is not required.
The background mesh is an octree grid with cubic cells.  %The mesh can be easily refined or coarsened locally  based on different adaptivity criteria.
 As an example of an application, we consider the modeling of contaminant transport in fractured porous media.
 One standard model leads to a coupled system of  advection--diffusion equations in a bulk (matrix) and along a surface (fracture). A series of numerical experiments with  both steady and unsteady problems and different embedded geometries illustrate the numerical properties of the hybrid approach.
The method demonstrates great flexibility in handling curvilinear or branching lower dimensional embedded structures.
 \end{abstract}


%\begin{AMS}
%  58J32, 65N15, 65N30, 76D45, 76T99
%\end{AMS}

%=============================================================================
%   Introduction
%=============================================================================

\section{Introduction}
Systems of coupled bulk--surface partial differential equations arise in many engineering and natural science applications. Examples include multiphase fluid dynamics with soluble or insoluble surfactants~\cite{GrossReuskenBook}, dynamics of biomembranes~\cite{bonito2011dynamics}, crystal growth~\cite{Crystal},  signaling in biological networks~\cite{Cells}, and transport of solute in fractured porous media \cite{alboin2002modeling}. In these and other applications, partial differential equations defined in a volume domain are coupled to another PDEs posed on a surface. The surface may be embedded in the bulk or belong to a boundary of the volume domain.

Recently, there has been a growing interest in developing methods for the numerical treatment of bulk--surface coupled PDEs.
Different approaches  can be distinguished depending on how the surface is recovered and equations are treated.
If a  tessellation of the volume into tetrahedra is available that fits the surface, then it is natural to introduce finite element spaces in the volume and on the induced triangulation of the surface. The resulting \textit{fitted} bulk--surface finite element method was studied  for the stationary bulk--surface advection--diffusion equations \cite{ER2013}, for non-linear reaction--diffusion systems modelling biological pattern formation  \cite{madzvamuse2016bulk,madzvamuse2015stability}, for the equations of the two--phase flow with surfactants~\cite{barrett2015stable,barrett2015stable2},  Darcy and transport--diffusion  equations in fractured porous media~\cite{alboin2002modeling}.  %application of the fitted FE method to a model of cell migration and chemotaxis can be found in~\cite{}.
%Other applications of the fitted FEM in \cite{Cells,hellander2012coupled,egger2015finite,elliott2015coupled,macdonald2016computational}.

\textit{Unfitted} finite element methods allow the surface to cut through the background tetrahedral mesh. In the class of finite element methods also known as cutFEM, Nitsche-XFEM or TraceFEM, standard background finite element spaces are employed, while the integration is performed over cut domains and over the embedded surface~ \cite{cutFEM,TraceFEM}. Additional stabilization terms are often added to ensure the robustness of the method with respect to small cut elements. The advantages of the unfitted approach are the efficiency in handling implicitly defined surfaces, complex geometries,   and  the flexibility in dealing with evolving domains. In the context of bulk--surface coupled problems,  cut finite element methods were recently applied to treat  stationary bulk--surface advection--diffusion equations \cite{gross2015trace}, %two-phase Stokes flow with soluble surfactants   \cite{hansbo2015cut},
coupled bulk-surface problems on time-dependent domains \cite{hansbo2016cut}, coupled elasticity problems \cite{cenanovic2015cut}. The hybrid  method developed in this paper belongs to the general class of unfitted methods and resembles the TraceFEM in how the surface PDE is treated.

The methods discussed above treat surfaces and interfaces sharply, i.e. as lower-dimensional manifolds. In the present paper we also  consider sharp interfaces. For the application of phase--field or other diffuse-interface approaches for coupled bulk--surface PDEs see,  for example, \cite{chen2014conservative,levine2005membrane,teigen2009diffuse}.

If the finite element method is a discretization of choice for the bulk problem, then it is natural to consider a finite element method for surface PDE as well. However,  depending on application, desired conservation properties, available software or personal experience, other discretizations such as finite volume or finite difference methods can be preferred for the PDE posed in the volume.
One possibility to reuse the same mesh for the surface PDE is to consider  a diffuse-interface approach. Alternatively, instead of smearing the interface, one may extend the PDE off the  surface to a narrow band containing the surface
in such a way that the restriction of the extended PDE solution back to the (sharp) surface solves the original equation on this surface. Further a conventional discretization is built for the resulting volume PDE in the narrow band~ \cite{BCOS01,olshanskii2016narrow}. The methods based on such extensions, however, increase the number of the active degrees of freedom for the discrete surface problem, may lead to degenerated PDE, need numerical boundary conditions and require smooth surfaces with no  geometrical singularities.

The present  paper develops a numerical method based on the sharp-interface representation, which uses a finite volume method to discretize the bulk PDE. Our goal is (i) to allow the surface to overlap with the background mesh in an arbitrary way, (ii) to avoid {building regular surface triangulation}, (iii)  to avoid any extension of the surface PDE to the bulk domain.  To accomplish these goals, we combine the monotone (i.e. satisfying the discrete maximum principle) finite volume method on  general meshes  \cite{Lipnikov:12,ChernyshenkoFV7:14}   with the trace finite element method on octree meshes from \cite{chernyshenko2015adaptive}.
In the octree TraceFEM one considers the bulk finite element space of piecewise trilinear globally continuous functions
and further uses the restrictions (traces) of these functions to the surface. These traces are further used in a variational formulation of the surface PDE. Effectively, this results in the integration of the standard polynomial functions over the (reconstructed) surface.  Only degrees of freedom from the cubic cells cut by the
surface are active for the surface problem. Surface parametrization is not required, no surface mesh is built,  no PDE extension off the surface is needed.  We shall see that the resulting hybrid FV--FE method is very robust with respect to the position of surfaces against the background mesh and is well suited for handling non-smooth surfaces and surfaces given implicitly.

One application of interest is the numerical simulation of the contaminant  transport and diffusion in fractured porous media. In this application, transport and diffusion along fractures are often modeled by PDEs posed on a set of piecewise-smooth surfaces, see, e.g.,~\cite{alboin2002modeling,fumagalli2013reduced,maryvska2005numerical,therrien1996three}; {see also \cite{alboin2002modeling,frac1,frac2,frac3} for a similar dimension reduction approach in simulation of flow in fractured porous media}.
Monotone (satisfying the DMP) finite volume methods on general meshes is the appealing tool for the solution of
equations for solute concentration in the porous matrix, see, e.g., ~\cite{ChernyshenkoFV7:14,Droniou:11,GaoWu:13,KapyrinFV7:14,LePotier:08,Lipnikov:12,ShengYuan:11}
(further references can be found in \cite{Droniou:14,FVCA7:14}).
 %Monotone, positivity preserving finite volume methods on polyhedral meshes is the appealing tool for solving the Darcy  equations~\cite{Nikitin:14,Terekhov:13} and equations for solute concentration in the porous matrix~\cite{KapyrinFV7:14,Kapyrin:08}.
 However, a straightforward application of this technique to model transport and diffusion along a fracture would require fitting the mesh or triangulating  the surface. For a large and complex net of fractures cutting through the porous matrix {this is a difficult task \cite{DPRF}, and} an efficient method avoids mesh fitting and surface triangulations. {
 Recently, extended finite element method  approximations have been extensively studied in transport and flow problems in fractured porous media, see the review \cite{flemisch2016review} and references therein.  In XFEM, one also avoids fitting of the background mesh to a fracture, but a separate mesh  is still required to represent the fracture.
 Besides the use of FV for the matrix problem, the approach in the present paper differs from those found in existing XFEM literature in the way the surface problem is discretized.}

While the present technique can be applied for tetrahedral or more general polyhedral  tessellations of the bulk domain, we consider octree grid with cubic cells here. This choice is not \textit{ad hoc}. Indeed, the Cartesian structure and built-in hierarchy of octree grids makes mesh adaptation, reconstruction and data access fast and easy. For these reasons, octree meshes became a common tool in  %image processing, the visualization of amorphous medium, computational free surface and multi-phase flows, and other applications where non-trivial geometries occur, see, for example, \cite{Losasso:04,Meagher,olshanskii2013octree,Pop09,Strain,Szeliski},
in computational mechanics and  several octree-based solvers  are available in the open source scientific computing software, \cite{DEAL2,Pop03}. However, an octree grid provides {at most} the first order (staircase) approximation of a general geometry. Allowing the surface to cut through the octree grid in an arbitrary way overcomes this issue, but challenges us with the problem of building efficient bulk--surface discretizations.   This paper demonstrates that the hybrid TraceFEM -- non-linear FV method complements the advantages of using octree grids by delivering {more accurate treatment of the surface PDE problem}. %The approach should be feasible for more general

The remainder of the paper is organized as follows. In section~\ref{s_setup} we recall the system of differential equations, boundary and interface conditions, which models the coupled bulk--interface (or ``matrix--fracture'' in the context of flows in porous media) advection--diffusion problem. Section~\ref{s_hybrid} gives the details of the hybrid discretization.
After laying out the main ideas behind the method, we discuss the non-linear monotone FV method for the bulk and the TraceFEM for
the surface equations, and further we introduce the required coupling.  Section~\ref{s_numer} presents the results of several numerical experiments with steady analytical solutions on  smooth and piecewise smooth branching surface. We also show the results of numerical simulation of the propagating front of solute concentration through fractured porous media.  %Some  remarks and conclusions are collected in closing section~\ref{s_concl}.

%Simulations of membrane-bound Turing patterns formation using the phase-field method \cite{levine2005membrane}
%Numerical methods based on diffuse-interface approach were developed in~\cite{teigen2009diffuse,chen2014conservative} and applied to simulate interfacial flows with soluble surfactants.

\section{Mathematical model}\label{s_setup}
In this section we recall the mathematical model of the contaminant  diffusion and transport in fractured  porous media. %We shall apply the hybrid FV-FE method to treat this problem numerically.
Assume the given bulk  domain $\Omega \subset \mathbb{R}^3$ and a piecewise smooth surface $\Gamma\subset\Omega$. The surface $\Gamma$ may have several connected components. If  $\Gamma$ has a boundary, we assume that $\dG\subset\dO$. %The situation, when $\Gamma$ has all of its boundary or a part of it inside $\Omega$ can be also handled by the present method, but requires a few extra technical detail, which are outlined in Remark~\ref{rem_G}.
Thus, we have the subdivision $\overline{\Omega}=\cup_{i=1,\dots,N}\overline{\Omega}_i$ into simply connected subdomains $\Omega_i$ such that $\overline{\Omega}_i\cap\overline{\Omega}_j\subset\Gamma$, $i\neq j$.

In each  $\Omega_i$, we assume a given Darcy velocity field of the fluid $\bw_i(\bx)$, $\bx \in \Omega_i$.
By $\bw_\Gamma(\bx)$, $\bx \in \rev{\Gamma}$, we denote the velocity field tangential to $\Gamma$ having the physical meaning of the flow rate through the cross-section of the fracture.  Consider an agent that is soluble in the fluid and transported by the flow in the bulk and along the fractures. The fractures are modeled by the surface $\Gamma$.
The solute \emph{volume} concentration (i.e., the one in the bulk domain $\Omega$) is denoted by $u$, $u_i=u|_{\Omega_i}$. The solute \emph{surface} concentration along $\Gamma$ is denoted by $v$.
Change of the concentration happens due to convection by the  velocity fields $\bw_i$ and $\bw_\Gamma$, diffusive fluxes in $\Omega_i$, diffusive flux on $\Gamma$, as well as the fluid exchange and diffusion flux between the fractures and the porous matrix.
These coupled processes can be modeled by the following system of equations~\cite{alboin2002modeling}, in subdomains,
\begin{equation} \label{diffeq1}
\left\{
\begin{aligned}
 \phi_i\frac{\partial u_i}{\partial t} + \Div(\bw_i  u_i- D_i \nabla u_i) &= f_i \quad \text{in}~ \Omega_i,\\
 u_i&=v \quad \text{on}~ \dO_i\cap\Gamma,
 \end{aligned}
\right.
 \end{equation}
and on the surface,
\begin{equation} \label{diffeq2}
 \phi_\Gamma\frac{\partial v}{\partial t} + \Div_\Gamma (\bw_\Gamma v - d D_\Gamma \nabla_\Gamma v) =  F_\Gamma(u)+ f_\Gamma  \quad \text{on}~~\Gamma,
 \end{equation}
where we employ the following notations:  $\nabla_\Gamma$,  $\Div_\Gamma$  denote the surface tangential gradient and divergence
 \begin{wrapfigure}{r}{0.4\textwidth}
\vspace{-20pt}
  \begin{center}
    \includegraphics[width=0.4\textwidth]{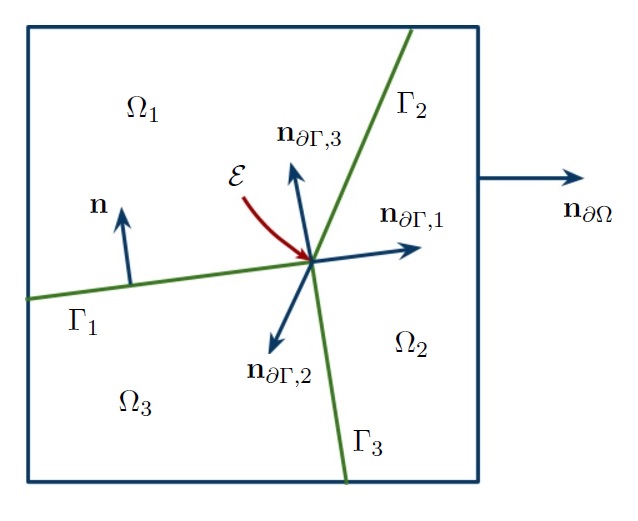}%{./Pictures/mainscreen1.png}
    \caption{2D illustration of our notation for a domain with triple fraction. %, $\Gamma=\cup_{i=1}^3\Gamma_i$  and $\mathcal{E}=\cap_{i=1}^3\overline{\Gamma}_i$.
    }
    \label{fig:Schema}
  \end{center}
  \vspace{-20pt}
  \vspace{1pt}
\end{wrapfigure}
operators; $F_\Gamma(u)$ stands for the  net flux of the solute per surface area due to fluid leakage and hydrodynamic dispersion;   $f_i$ and $f_\Gamma$ are given source terms in the subdomains and in the fracture;  $D_i$ denotes the diffusion tensor in the porous matrix; the surface diffusion tensor is $D_\Gamma$.
Both $D_i$, $i=1,\dots,N$, and $D_\Gamma$ are symmetric and positive definite; $d>0$ is the fracture width coefficient; $\phi_i>0$ and $\phi_\Gamma>0$ are the constant porosity coefficients for the bulk and the fracture.

The total surface flux $F_\Gamma(u)$ represents the contribution of the bulk to the solute transport in the fracture. The mass balance at  $\Gamma$ leads to the equation
\begin{equation}\label{Flux}
 F_\Gamma(u)= [-D \bn \cdot \nabla u + (\bn \cdot\bw)u]_\Gamma,
\end{equation}
where $\bn$ is a unit normal vector at $\Gamma$,
$[w(\bx)]_\Gamma = \lim\limits_{\eps\to0} w(\bx-\eps\bn)-\lim\limits_{\eps\to0} w(\bx+\eps\bn)$, $\bx\in\Gamma$,
denotes the jump of $w$ across $\Gamma$ in the direction of $\bn$.

If $\Gamma$ is  \textit{piecewise} smooth, then we need further conditions on the edges.
Assume an edge $\mathcal{E}$ is shared by  $M$  smooth components $\Gamma_i\subset\Gamma$.
Let $v_{j}=v$ on $\Gamma_j$, while $\bw_{\Gamma,j}=\bw_{\Gamma}$, $d_j=d$, $D_{\Gamma,j}=D_\Gamma$ on $\Gamma_j$, and $\bn_{\dG,j}$ is the outward  normal vector to $\dG_j$ in the plane tangential to $\Gamma_j$, cf. Figure~\ref{fig:Schema}.
The conservation of fluid mass yields
\begin{equation}\label{cont_w}
\sum_{j=1}^M\bw_{\Gamma,j}\cdot\bn_{\dG,j}=0\quad \text{on}~~\mathcal{E}.
\end{equation}
We assume the continuity of concentration over $\mathcal{E}$,
\begin{equation}\label{cont_v}
v_{1}=\dots=v_{M}\quad \text{on}~~\mathcal{E}.
\end{equation}
We also assume the conservation of solute flux over the edge. Thanks to \eqref{cont_w} and \eqref{cont_v}, this yields the condition:
\begin{equation}\label{cont_F}
\sum_{j=1}^M d_j(D_{\Gamma,j}\bn_{\dG,j})\cdot\nabla_\Gamma v_j=0\quad \text{on}~~\mathcal{E}.
\end{equation}

Finally, we prescribe Dirichlet's boundary conditions for the concentration $u$ and $v$ on $\dO_D$ and $\dG_D$ and homogeneous Neumann's boundary conditions on $\dO_N$ and $\dG_N$, respectively, with $\overline{\dO}=
\overline{\dO_D}\cup\overline{\dO_N}$ and $\overline{\dG}=\overline{\dG_D}\cup\overline{\dG_N}$. Initial conditions are given by the known concentration $u_0$ and $v_0$ at $t=0$. We have
\begin{equation} \label{bc}
\left\{
\begin{aligned}
 D_i \bn_\dO \cdot \nabla u&=0 \quad \text{on}~ \dO_N,\\
 u&=u_D \quad \text{on}~ \dO_D,\\
 u|_{t=0}&=u_0 \quad \text{in}~ \Omega,\\
 \end{aligned}
\right.\qquad \left\{
\begin{aligned}
D_{\Gamma}\bn_{\dG}\cdot\nabla_\Gamma v&=0 \quad \text{on}~ \dG_N,\\
 v&=v_D \quad \text{on}~ \dG_D,\\
 v|_{t=0}&=v_0 \quad \text{on}~~\Gamma.
 \end{aligned}
\right.
 \end{equation}

\begin{remark}\rm Bulk--surface coupled systems of advection-diffusion PDEs appear in different applications,
e.g.  in multiphase fluid dynamics \cite{GrossReuskenBook} and biological applications \cite{bonito2011dynamics}. In these and other models, the continuity of the concentration over the embedded surface (second equation in \eqref{diffeq1}) may {be replaced by} another suitable   constitutive equation for modeling of the surface adsorption/desorption.
For fluid--fluid interfaces or biological membranes, one often assumes that the surface passively evolves with the flow, and hence there is no contribution of the advective flux to the total flux $F_\Gamma(u)$ on $\Gamma$. A standard model for the diffusive flux between the surface and the bulk, cf.~\cite{Ravera}, is as follows:
\begin{equation} \label{eq4}
-D_i \bn \cdot \nabla u_i  =  k_{i,a} g_i(v) u_i - k_{i,d} f_i(v),\quad\text{on}~\Gamma,
\end{equation}
with $k_{i,a}$, $k_{i,d}$ positive adsorption and desorption coefficients that describe the kinetics.  Basic choices for $g$, $f$ are the following:
\[
  g(v)=1, \quad f(v)=v \quad \text{(Henry)}\quad \text{or}\quad g(v)=1- \frac{v}{v_{\infty}},\quad f(v)=v \quad \text{(Langmuir)},
\]
where $v_\infty$ is a constant that quantifies the maximal concentration on $\Gamma$. Further options are given in \cite{Ravera}.
Often in literature on the two-pase flows the Robin condition in \eqref{eq4} is replaced by the ``instantaneous'' adsorption and desorption condition
\begin{equation} \label{eq4a}
k_{i,a} g_i(v) u_i = k_{i,d} f_i(v),\quad\text{on}~\Gamma,
\end{equation}
These interface conditions can be also handled through obvious modifications  of our numerical method. {We include one numerical example with \eqref{eq4a} and Henry law in Section~\ref{s_numer}.} At the same time, treating evolving interfaces needs more developments and is not considered here.
\end{remark}

\section{Hybrid finite volume -- finite element method}\label{s_hybrid}

\subsection{Summary of the method}

Assume a Cartesian background mesh with cubic cells. We allow local refinement of the mesh by sequential division
of any cubic cell into 8 cubic subcells. This leads to a grid with an octree hierarchical  structure. This mesh gives the tessellation   $\T_h$ of the computational domain $\Omega$, $\overline{\Omega}=\cup_{T\in\T_h} \overline{T}$.
The surface  $\Gamma\subset\Omega$ cuts through the mesh in an arbitrary way. For the \textit{purpose of numerical integration},
instead of $\Gamma$ we consider $\Gamma_h$, a given polygonal approximation of $\Gamma$. If $\Gamma$ has a curvature, then $\Gamma_h$ is reconstructed as a second order approximation of $\Gamma$. We shall describe the reconstruction algorithm further in the section. We assume that similar to $\Gamma$, the reconstructed surface $\Gamma_h$ divides $\Omega$ into $N$ subdomains $\Omega_{i,h}$, and $\dG_h\subset\dO$. We do not {assume} any restrictions on how $\Gamma_h$ intersects the background mesh.

The induced tessellation of  $\Omega_{i,h}$ can be considered as a subdivision of the volume into general polyhedra.
Hence, for the transport and diffusion in the matrix we apply  a non-linear FV method devised on general polyhedral meshes in \cite{Lipnikov:12,ChernyshenkoFV7:14}, which is monotone and has compact stencil. The trace of the background mesh on $\Gamma_h$ induces {a} `triangulation' of the fracture, which is very irregular, and so we do not use it \rev{to} build a discretization method.
To handle transport and diffusion along the fracture, we first consider finite element space of piecewise trilinear functions
for the \textit{volume} octree mesh $\T_h$. We further, formally, consider the restrictions (traces) of these background functions on $\Gamma_h$ and use them in a finite element integral  form over $\Gamma_h$.
Thus the irregular triangulation of $\Gamma_h$ is used \textit{for numerical integration only}, while the trial and test functions are tailored to the background \textit{regular} mesh. {Available analysis and numerical experience suggest} that the {approximation and convergence} properties of this trace finite element method
{depend only on the mesh size and refinement strategy for}  the background mesh, and they are independent on how $\Gamma_h$ intersects $\T_h$. The TraceFEM was devised and first analysed in \cite{ORG09} and extended for the octree meshes in \cite{chernyshenko2015adaptive}.  A natural way to couple two approaches is to use the
restriction of the background FE solution on $\Gamma_h$ as the boundary data for the FV method and to compute the FV two-side fluxes on $\Gamma_h$ to provide the source terms for the surface discrete equation. We provide details of each of these steps in sections~\ref{s_FV}--\ref{s3} below.

\subsection{Reconstructed surface}\label{s_rec}
The reconstructed surface $\Gamma_h$ is a $C^{0,1}$ surface that can be partitioned in planar triangular segments:
\begin{equation} \label{defgammah}
 \Gamma_h=\bigcup\limits_{K\in\mathcal{F}_h} K,
\end{equation}
where $\mathcal{F}_h$ is the set of all  triangular segments $K$.
Without loss of generality we assume that for any $K\in\mathcal{F}_h$ there is only \textit{one}
cell $T_K\in\mathcal{T}_h$ such that $K\subset T_K$ (if $K$ lies on a face shared by two cells, any of these
two cells can be chosen as $T_K$).

In practice, we construct $\Gamma_h$ as follows. For each connected piece of $\Gamma$ let $\phi$ be a Lipschitz-continuous level set function, such that $\phi(\bx)=0$ on $\Gamma$. We set $\phi_h=I(\phi)$  a nodal interpolant of $\phi$ by a  piecewise trilinear continuous function with respect
to the octree grid $\mathcal{T}_h$.  Consider the zero level set of $\phi_h$,
\[
\widetilde{\Gamma}_h:=\{\bx\in\Omega\,:\, \phi_h(\bx)=0 \}.
\]
If $\Gamma$  is smooth, then  $\widetilde{\Gamma}_h$ is an approximation to $\Gamma$ in the following sence:
\begin{equation}\label{eq_dist}
\mbox{dist}(\Gamma,{\widetilde{\Gamma}_h})\le ch^2_{\rm loc},\qquad |\bn(\bx)-\bn_h(\tilde\bx)|\le c h_{\rm loc},
\end{equation}
where $\bx$ is the closest point on $\Gamma$ for $\tilde\bx\in\widetilde{\Gamma}_h$ and $h_{\rm loc}$ is the local mesh size.
We note that in some applications, $\phi_h$ is computed from a solution of a discrete indicator function equation (e.g., in the level set or the volume of fluid methods), without any direct knowledge of $\Gamma$.

\begin{figure}
%\vspace{-20pt}
  \begin{center}a)
    \includegraphics[width=0.31\textwidth]{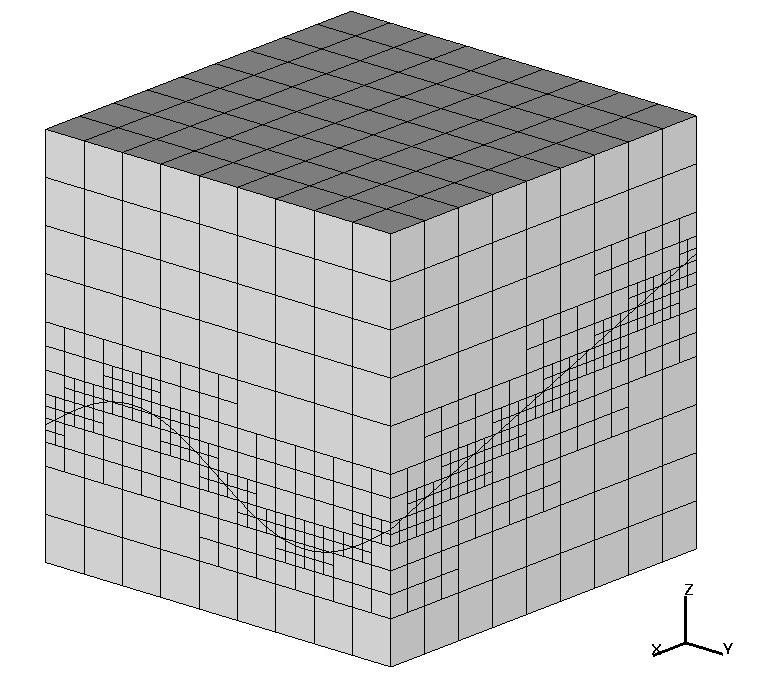}b)
    \includegraphics[width=0.31\textwidth]{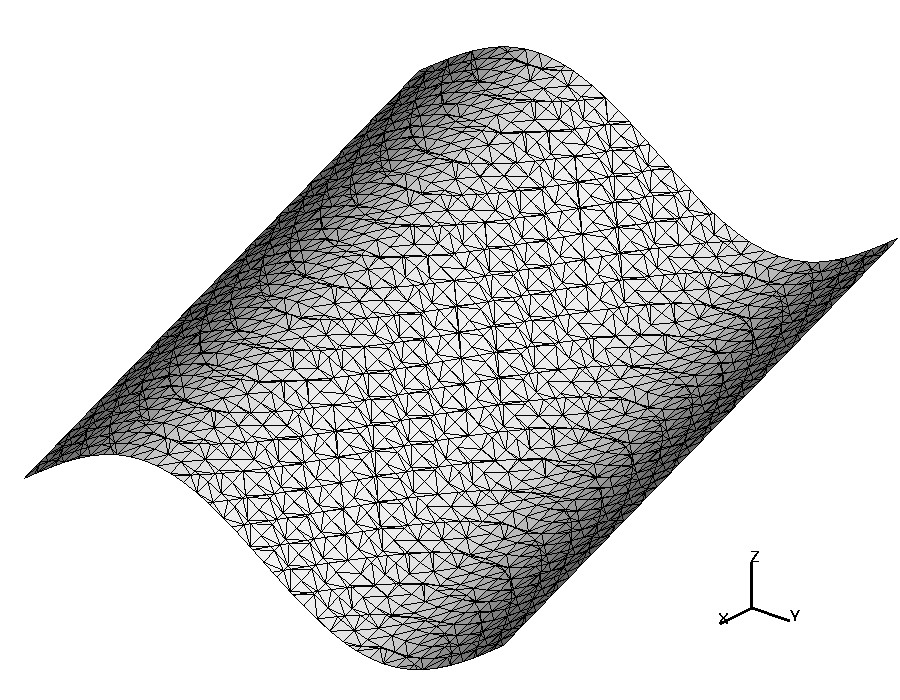} c)
    \includegraphics[width=0.29\textwidth]{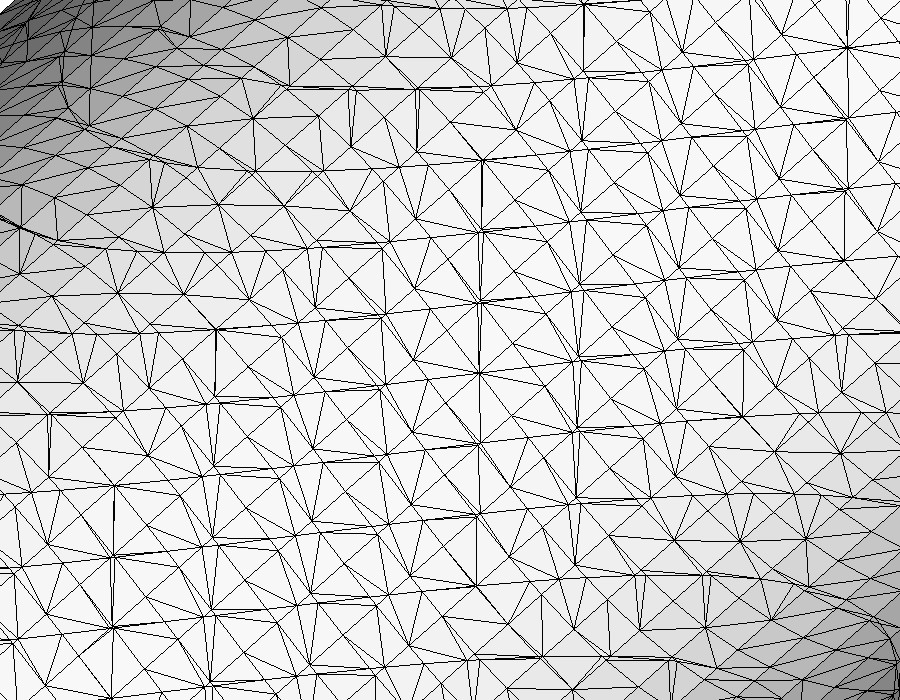}
    \caption{a) Example of a  bulk domain with a fracture. In this example, the background mesh is refined near the fracture; b) The reconstructed $\Gamma_h$; c) The zoom-in of the induced surface triangulation $\F_h$.}
    \label{fig1}
  \end{center}
 % \vspace{-20pt}
%  \vspace{1pt}
\end{figure}
{Note that $\widetilde{\Gamma}_h$ is still not completely suitable for our purposes, since $\phi_h$ is \emph{trilinear}
and so numerical integration over its zero level is not straightforward. Therefore, we next build a suitable polygonal approximation of $\widetilde{\Gamma}_h$ which is our final $\Gamma_h$.}
Once $\phi_h$ is computed, we recover $\Gamma_h$ by the cubical marching squares method from \cite{MCM2} (a variant of the very well-known marching cubes method). The method provides a triangulation of $\widetilde{\Gamma}_h$ within each
cube such that {$\Gamma_h$ is  continuous over cubes interfaces}, the number of triangles within each cube is finite and bounded
by a constant independent of $\widetilde{\Gamma}_h$ and a number of refinement levels. Moreover, the vertices of triangles from $\mathcal{F}_h$ are lying on $\widetilde{\Gamma}_h$. This final discrete surface $\Gamma_h$ is still an approximation of $\Gamma$ in the sense of \eqref{eq_dist}. A example of bulk domain with embedded surface and background mesh is illustrated in Figure~\ref{fig1}.

Note that the resulting triangulation $\mathcal{F}_h$  is \emph{not} necessarily  regular, i.e. elements from $T$ may have very small internal angles and the size of neighboring triangles can vary strongly. Thus, $\Gamma_h$ is not { a regular triangulation of $\Gamma$.} The surface triangulation $\mathcal{F}_h$ is used only to define quadratures in the finite element method, while approximation properties of the method depend on the volumetric octree mesh.

\subsection{Monotone finite volume method} \label{s_FV}
First we consider a FV method for the advection-diffusion equation \eqref{diffeq1} in each subdomain $\Omega_{i,h}$.
Let $\T_{i,h}$ be the tessellation  of $\Omega_{i,h}$ into non-intersected polyhedra, which is induced by overlapping $\Omega_{i,h}$ and the background mesh $\T_h$.   Since the background mesh is the octree Cartesian, each element $T\in\T_{i,h}$ is either a cube, if it lies in the interior of $\Omega_{i,h}$, or a cut cube, if $\dO_{i,h}$ intersects a background cell from $\T_h$. We assume the octree grid is gradely refined, i.e. the sizes of two neighbouring elements of $\T_h$ can differ at most by a factor of two.  Such octree grids are also known as balanced. The method applies for unbalanced octrees, but in our experiments we use balanced grids.
For the balanced grid,  each interior cell may have from 6 to 24 neighboring cells (cells sharing a face). In the FV method we treat such cells as a polyhedra with up to 24 faces.   Since we reconstruct $\Gamma$ inside each cell as a triangulated surface without holes, the cut cell from $\T_{i,h}$ can be treated as a general polyhedral element as well. By $\F_{i,h}$ we denote the set of all faces of polyhedra from $\T_{i,h}$.

The FV discretization below is applied to each subdomain $\Omega_i$ separately, so we will skip in this section the  redundant index $i$ for the concentration, coefficients and the flow vector field in $\Omega_i$. Note that $\dO_{i,D}$
includes the fracture part of the boundary of $\dO_i$.

As the first step,  we assume a time discretization (say, the implicit Euler method) and consider the mixed form of  \eqref{diffeq1}  and  boundary conditions
\begin{equation}\label{lap}
\begin{array}{rclll}
  \bq = \bw u - D \nabla u, \quad
  \tilde\phi\,u+\mbox{\rm div } \bq  &=& f \quad     & \mbox{~in} & \Omega_i,  \\
       u &=& \tilde{u}_D                     & \mbox{~on} & \dO_{i,D} ,  \\
       -D \bn_\dO \cdot \nabla u &=&0 & \text{~on}& \dO_{i,N},
\end{array}
\end{equation}
where the right hand side $f$ accounts for the source term and for the values of concentration from the previous time step, $\tilde \phi$ \rev{is the porosity coefficient scaled by the reciprocal of the time step size}, and $\dO_{i,D}$ includes $\dO_i\cap\Gamma$, where $\tilde{u}_D=v$.
%Here $D$ is  the diffusion tensor which can be the scalar coefficient  $D=D_i\bI$ in $\Omega_i$ where $\bI$ is the  identity tensor and $D_i$ is the diffusion scalar coefficient from \eqref{diffeq1}.
%In this paper, we consider the case of isotropic diffusion, although the FV method below is applicable to  symmetric positive definite tensors $D$.

For a cell $T\in\T_{i,h}$, $\bx_{T}$ denotes the barycenter of $T$, and $u_T$ denotes the averaged concentration. We formally assign $u_T$ to $\bx_T$.
Integrating the mass balance equation \eqref{lap} over $T$ and using the divergence theorem, we obtain:
\begin{equation}\label{green2}
 \tilde\phi|T|u_T +\sum\limits_{\F \in \partial T} \bq_\F\cdot \bn_\F  =  \int_T f \,{\rm d} x,
  \qquad
  \bq_{\F} = \frac{1}{|\F|} \int_{\F}  \bq \,{\rm d} s,
\end{equation}
where $\bq_{\F} \cdot \bn_\F$ is the averaged normal flux across face $\F$,  and $\bn_\F$  is the normal vector
on $\F$ pointing outward for $T$; $|\F|$ ($|T|$) denotes the area (volume) of $\F$ ($T$).
The Dirichlet boundary data on faces $\F\in \partial\Omega_D$ will be accounted in the scheme via boundary faces concentration values
$u_\F = \frac{1}{|\F|}\int_{\F}  u_D \,{\rm d} s$. We assume that $u_\F$ are assigned to barycenters of faces.
 Enforcing homogeneous Neumann boundary conditions on faces from $\dO_{i,N}$ is straightforward, for $\F\in  \partial T\cap\dO_{N}$ the normal flux $\bq_\F\cdot \bn_\F$ in \eqref{green2} is  set to 0.

In the conventional cell-centered FV method, the normal flux $\bq_\F\cdot \bn_\F$ is replaced by its discrete counterpart $\bq_{\F,h} \cdot \bn_\F$, which is computed
from cell concentrations $u_T$ and boundary data $u_\F$. For simplicity of presentation we shall omit subscript $h$ in notations of the discrete flux.
The discrete flux is the combination of the diffusive and convective fluxes and we discretize them separately
following  \cite{Lipnikov:10,Lipnikov:12,NikitinVassilevski:10}.

\begin{figure}[h!]
\centering
\includegraphics[width=.45\textwidth]{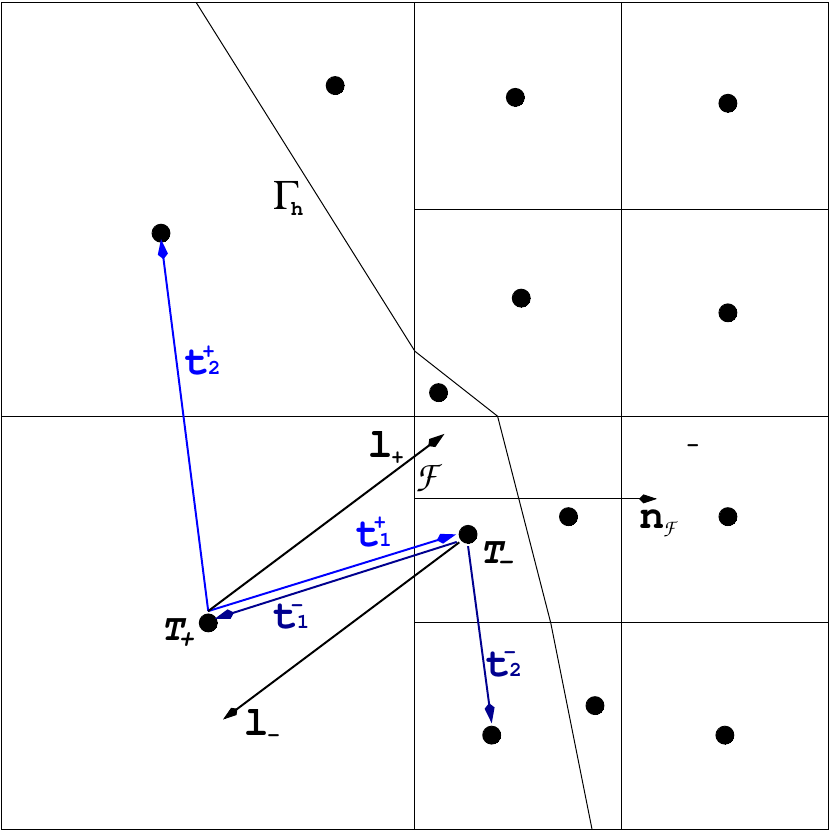}
\caption{For the 2D case, the figure illustrates our constructions for the approximation of the directional derivative $\bl_{\F}\cdot\nabla u$ on the face (edge in 2D) $\F=\overline{T_+}\cap\overline{T_-}$. Bold dots
show the barycenters $\bx_T$ of the cells from $\T_{1,h}$ and $\T_{2,h}$. The mean values of the concentration are assigned to these barycenters.}
\label{fig:dupl}
\end{figure}

For $T\in\T_{i,h}$, we define $\omega(T):=\{T'\in\T_{i,h}\,|\,\mbox{area}(\overline{T}'\cap \overline{T})\neq 0\}$, the set of all neighboring cells of $T$, and
 $\omega_{\partial}(T):=\{F\in\F_{i,h}\,|\,F{\subset} \partial T\cap \partial\Omega_{i,D}\}$, the set all faces of $T$ with prescribed Dirichlet data.  For $T\in\T_{i,h}$, the set of points $\mathcal{P}$ collects all barycenters of the elements from $\omega(T)$ and $\omega_{\partial}(T)$.  Furthermore, for each $T\in\T_{i,h}$ we define the bundle of vectors, $\bv(T):=\{\bt\in\mathbb{R}^3\,|\,\bt=\by-\bx_T,~\by\in \mathcal{P}(T)\}$.

Consider an arbitrary internal face $\F$ shared by two cells $T_+, T_-$ from $\T_{i,h}$ and assume that $\bn_\F$ points from $T_+$ to $T_-$. We introduce the co-normal vector $\bl_{\F}= D \bn_\F$.
Vector $\bl_{\F}$  can make a nonzero angle with $\bn_\F$ in the case of an anisotropic diffusion tensor.
To define the discrete {\em diffusive} flux on $\F$, we first %(see Fig.~\ref{fig:dupl} for a 2D example and notations),
consider three vectors $\bt_{i}^+\in \bv(T_+)$, $i=1,2,3$,
such that for the co-normal vector $\bl_{\F} = D \bn_f$ we have
\begin{equation}\label{ab}
  \bl_{\F} =
  \alpha_+~ \bt_{1}^+ + \beta_+~ \bt_{2}^+ + \gamma_+~ \bt_{3}^+,
\end{equation}
with non-negative coefficients $\alpha_+$, $\beta_+$ and $\gamma_+$. Such a triplet can be always found,  (in some rare pathological situations, one has to expand $\mathcal{P}(T_+)$ slightly, cf. \cite{DanilovVassilevski:09}).

The normal flux is the directional derivative along the co-normal vector $\bl_+ := \bl_{\F}$, and hence
it can also be represented as the linear combination of three derivatives along $\bt_{i}^+$. The latter  are approximated by central differences (may reduce to one side differences near Dirichlet boundaries). Thus, we get
$\bq_\F\cdot\bn_F\approx q_+$,
\begin{equation}\label{ab2}
 q_+=
  \alpha'_{+}~ (u_+ - u_{+,1}) + \beta'_{+}~ (u_+ - u_{+,2}) +
  \gamma'_{+}~ (u_+ - u_{+,3}),\qquad u_+=u(\bx_{T_+}),~ u_{+,i}=u(\bx_{T_+}+\bt_{i}^+),
\end{equation}
where coefficients $\alpha'_{+}, \beta'_{+}, \gamma'_{+}$ are computed from $\alpha$, $\beta$, $\gamma$ in  \eqref{ab}
for the cell $T_+$, using the simple  scaling with $|\bt_{i}^+| / |\bl_{\F}|$.
For the same co-normal vector one has  another decomposition based on $\bv(T_-)$ vector bundle,
$
\bl_- := -\bl_{\F} =
  \alpha_-~ \bt_{1}^- + \beta_-~ \bt_{2}^- + \gamma_-~ \bt_{3}^-$, $\bt_{i}^-\in \bv(T_-)$. This decomposition yields another approximation,  $\bq_\F\cdot\bn_F\approx q_-$:
\begin{equation}\label{ab3}
  q_- =
  \alpha'_{-}~ (u_- - u_{-,1}) + \beta'_{-}~ (u_- - u_{-,2}) +
  \gamma'_{-}~ (u_- - u_{-,3}),\qquad u_-=u(\bx_{T_-}),~ u_{-,i}=u(\bx_{T_-}+\bt_{i}^-),
\end{equation}
with non-negative coefficients $\alpha'_-$, $\beta'_-$ and $\gamma'_-$. Figure~\ref{fig:dupl} illustrates the construction in 2D.

Now we can take a linear combination of \eqref{ab2} and \eqref{ab3} with non-negative coefficients $\mu_+$ and $\mu_-$:
\begin{equation}\label{two-flux}
  \bq_{\F} \cdot \bn_{\F} = \mu_+ q_+ + \mu_- (- q_-).
\end{equation}
The discrete flux \eqref{two-flux}  approximates the differential one if   $\mu_+$, $\mu_-$  satisfy
\begin{equation}
\label{mupm1}
\mu_+ + \mu_- = 1.
\end{equation}
Following \cite{Lipnikov:12}, to construct the {\it monotone}  FV discretization, we set  both representations of the flux equal:
\begin{equation}
\label{fvdmp_mus}
  \mu_+ q_+ = - \mu_- q_-.
\end{equation}
If $| q_+| = |q_-| = 0$, then the solution of \eqref{mupm1},\eqref{fvdmp_mus} in not unique. In this case we choose $\mu_+ = \mu_- = 1/2$.
Otherwise, the solution is given by
\begin{equation*}
\mu_+ = \frac{q_-}{q_- - q_+}, \qquad \mu_- = \frac{q_+}{q_+ - q_-}.
\end{equation*}
If $q_+ q_- > 0$, we avoid   potentially degenerate case by applying the modification from~\cite{ShengYuan:11}; see also formulas in \cite{Lipnikov:12}, p.~374.
We note that the resulting multi-point flux approximation is nonlinear and compact, i.e. the stencil includes the values of concentration only from neighboring cells.

To define the normal component of the discrete \textit{advective} flux
$\bq_{\F,a} =\frac{1}{|\F|} \int_{\F}  u\bw \,{\rm d}s$,
we adopt the nonlinear  upwind approximation (subscript $h$ is again omitted for the sake of notation):
\begin{equation}\label{vRvR}
  \bq_{\F,a} \cdot \bn_{\F} = w_{\F}^+  \cR_{T^+}(\bx_{\F}) + w_{\F}^-  \cR_{T^-}(\bx_{\F}),
\end{equation}
where
$$
w_{\F}^+ =\frac12 (w_{\F}+ |w_{\F}|), \quad w_{\F}^- =\frac12 (w_{\F}- |w_{\F}|), \quad w_{\F} = \frac{1}{|\F|} \int_{\F}  \bw\cdot\bn_{\F} \,{\rm d} s,
$$
$\cR_T$ is a linear reconstruction of the concentration
over cell $T$ which depends on the concentration values from neighboring cells.

On each cell $T$, the linear reconstruction is defined by
\begin{equation}\label{reconst}
\cR_T (\bx) = \left\{ \begin{array}{ll} u_T + \cL_T \bg_T \cdot (\bx-\bx_{T}), & \bx\in T, \\
                                       0, & \bx\notin T, \end{array} \right.
\end{equation}
where $\bg_T$ denotes the gradient of the  linear reconstruction of concentration in $\bx_T$, and $\cL_T$ is a slope limiting operator. The gradient is recovered from the best affine least-square fit for $u_h$ over a subset of barycenter nodes and, possibly, the boundary data nodes from cells neighboring $T$.
The slope limiting operator $\cL_T$ is introduced to avoid non-physical extrema. It provides the smallest possible changes of the reconstructed least-square slope. Details can be found in  \cite{Lipnikov:10,Lipnikov:12,NikitinVassilevski:10}.

Replacing fluxes in equations \eqref{green2} by their numerical approximations, we
obtain a system of nonlinear equations
\begin{equation} \label{sol:system}
  \Phi\bU+ \bM(\bU)\, \bU = \bF(\bU),
  \qquad
  \bM(\bU) = \bM_{dif}(\bU) + \bM_{adv}(\bU),
\end{equation}
with a diagonal matrix $\Phi$. For any fixed vector $\bV$, $\bM(\bV)$ is a square sparse matrix,  $\bF(\bV)$ is a right-hand side vector.
Matrix $\bM_{dif}$ is an M-matrix which has  diagonal dominance in rows.
The stencil of this matrix is compact, each row contains non-zero off-diagonal entries corresponding mainly (and in most cases only) to degrees of freedom at the cells sharing a face with the current cell.
 For a  cubic uniform mesh and the Poisson equation, the matrix $\bM_{dif}$ corresponds to the conventional seven-point stencil.
Although matrix $\bM_{adv}$ has no diagonal dominance in rows,
it can be shown, cf.~\cite{Lipnikov:12}, that the solution to \eqref{sol:system}
satisfies the discrete maximum principle.

\subsection{The trace finite element method} \label{s_FEM}

Consider now the volumetric finite element space of all piecewise trilinear continuous functions with respect to the bulk octree mesh $\mathcal{T}_h$:
\begin{equation}
 V_h:=\{v_h\in C(\Omega)\ |\ v|_{S}\in Q_1~~ \forall\ S \in\mathcal{T}_h\},\quad\text{with}~~
 Q_1=\mbox{span}\{1,x_1,x_2,x_3,x_1x_2,x_1x_3,x_2x_3,x_1x_2x_3\}.
 \label{e:2.6}
\end{equation}
The surface finite element space is \textit{the space of traces on $\Gamma_h$ of all piecewise trilinear continuous functions with respect to the outer triangulation $\mathcal{T}_h$}  defined as follows
\begin{equation}
 V_h^{\Gamma}:=\{\psi_h\in H^1(\Gamma_h)\ |\ \exists ~ v_h\in V_h\  \text{such that }\ \psi_h=v_h|_{\Gamma_h}\}.
\label{e:fem-space}
\end{equation}

Given the surface finite element space $V_h^{\Gamma}$,  the finite element
discretization of \eqref{diffeq2} is as follows:   Find $v_h\in V_h^{\Gamma}$ such that $v_h|_{\partial\Gamma_{D,h}}=v_D^h$ and
\begin{multline}
\int_{\Gamma_h}\left(  \phi_{\Gamma,h}\frac{\partial v_h}{\partial t}w_h +  d_h D_{\Gamma,h}\nath v_h\cdot\nath w_h\, +(\bw_h\cdot\nath v_h) w_h\right)\, + (\Div_{\Gamma_h}\bw_h) \, {w_h}v_h \ds_h\\ =\int_{\Gamma_h}( F_{\Gamma,h}(u_h)+ f_{\Gamma,h}) w_h\, \ds_h \label{FEM}
\end{multline}
for all $w_h\in V_h^{\Gamma}$, s.t. $w_h|_{\partial\Gamma_{D,h}}=0$ . Here $\mathbf{w}_h$, $v_D^h$, $d_h$, $D_{\Gamma,h}$ and $f_{\Gamma,h}$ are  the problem data lifted from $\Gamma$ to $\Gamma_h$, in the case if $\Gamma\neq\Gamma_h$. The bulk domain contributes through the flux $F_{\Gamma,h}(u_h)$, which is reconstructed from the numerical concentration in the porous matrix. %A well-posedness result for \eqref{FEM} will be proved in the next section.

Similar to the plain Galerkin finite element for  advection-diffusion equations the method \eqref{FEM} is prone to instability unless mesh is sufficiently fine such that the mesh Peclet number is less than one.
Following \cite{ORXimanum}, we consider the SUPG stabilized TraceFEM.  The stabilized formulation reads:
 Find $v_h\in V_h^{\Gamma}$ such that
\begin{multline}
 \int_{\Gamma_h}\left(  \phi_{\Gamma,h}\frac{\partial v_h}{\partial t}w_h +  d_h D_{\Gamma,h}\nath v_h\cdot\nath w_h\, +(\bw_h\cdot\nath v_h) w_h\right)\, + (\Div_{\Gamma_h}\bw_h) \, {w_h}v_h \ds_h \\
  +\sum_{{T}\in\mathcal{F}_h}\delta_K\int_{K}(\phi_{\Gamma,h}\frac{\partial v_h}{\partial t}-{d_h\Div_{\Gamma_h}D_{\Gamma,h}\nabla_{\Gamma_h}}v_h + \bw_h\cdot\nath v_h + (\Div_{\Gamma_h}\bw_h) \, v_h)\,\bw_h\cdot\nath w_h\, \ds_h\\ =\int_{\Gamma_h}( F_{\Gamma,h}(u_h)+ f_{\Gamma,h}) w_h\, \ds_h + \sum_{K\in\mathcal{F}_h}\delta_K\int_{K}( F_{\Gamma,h}(u_h)+ f_{\Gamma,h})(\bw_h\cdot\nath w_h)\, \ds_h\quad \forall~ w_h\in V_h^{\Gamma}. \label{FEM_SUPG}
\end{multline}
{For the definition of $K\in\mathcal{F}_h$, $T_K\in\T_h$ we refer to section~\ref{s_rec}.}
The stabilization parameter  $\delta_K$ depends on $K \subset T_K$. The side length of the cubic cell $T_K$ is denoted by $h_{T_K}$.  Let $\displaystyle \mathsf{Pe}_K:=\frac{h_{T_K} \|\mathbf{w}_h\|_{L^\infty(K)}}{2\eps}$
be the cell Peclet number.
We take
\begin{equation}
 {\delta_K}=
\left\{
\begin{aligned}
&\frac{\delta_0 h_{T_K}}{\|\mathbf{w}_h\|_{L^\infty(K)}} &&\quad \hbox{ if } \mathsf{Pe}_K> 1,\\
&\frac{\delta_1 h^2_{T_K}}{\eps}  &&\quad \hbox{ if } \mathsf{Pe}_K\leq 1,
\end{aligned}
\right. \label{e:2.10}
\end{equation}
with some given positive constants  $\delta_0,\delta_1\geq 0$.

For the matrix--vector representation of the TraceFEM one uses the nodal basis of the bulk finite element space $V_h$ rather than tries to construct a basis in $V_h^{\Gamma}$.  This convenient choice, however, has some consequences. In general, the restrictions to $\Gamma_h$ of the outer nodal basis functions on $\Gamma_h$ can be linear dependent or (in most cases) almost linear dependent.
This and small cuts of background cells lead to badly conditioned mass and stiffness matrices. In recent
years stabilizations have been developed which are easy to implement and result in
matrices with acceptable condition numbers, see the overview in \cite{TraceFEM}.
In this paper we use the ``full gradient'' stabilization of the TraceFEM~\cite{Alg2,Reusken2014}. In this variant of the method, one modifies the surface diffusion part of the  method~\eqref{FEM} to include the normal part of the gradient:
\[
\int_{\Gamma_h}d_h D_{\Gamma,h}\nath v_h\cdot\nath w_h\,\ds_h\quad\text{yields to}\quad
\int_{\Gamma_h}d_h D_{\Gamma,h}\nabla v_h\cdot\nabla w_h\,\ds_h.
\]
We note that the method remains consistent on smooth surfaces (up to second order geometric errors), since the true surface solution extended off the surface along normal directions satisfies both variational formulations on $\Gamma$.   The modification improves algebraic properties of the (diagonally scaled) stiffness matrix of the method~\cite{Reusken2014}. The full-gradient method  uses the background finite element space $V_h$ instead of the surface finite element space $V_h^\Gamma$ in \eqref{FEM}. However, practical implementation of both methods uses the frame of all bulk finite element nodal basis functions $\phi_i\in V_h$ such that  $\mbox{supp}(\phi_i)\cap\Gamma_h\neq\emptyset$. Hence the active degrees of freedom  in both methods are the same. The stiffness matrices are, however, different.

\subsection{Coupling between discrete bulk and surface equations}\label{s3}

The equations in the bulk and on the surface are coupled through the boundary condition $u_i=v$ on $\partial\Omega_{i,h}\cap\Gamma_h$ (second equation in \eqref{diffeq1}) and the net flux $F_{\Gamma_h}(u)$ on $\Gamma_h$, which stands as the source term in the surface equation \eqref{diffeq2}. On $\Gamma_h$ the solution $v_h$ is defined as a trace of the background finite element piecewise trilinear function. The averaged value
 of $v_h$ is computed on each surface triangle $K\in\F_h$ using a standard quadrature rule. These values assigned to the barycenters of $K$ from $\F_h$ serve as the Dirichlet boundary data for the FV method on $\Gamma_h$.
 The discrete diffusive and convective fluxes are assigned to barycenters of all faces on $\T_{i,h}$, $i=1,\dots,N$.
Since each triangle $K\in\F_h$ is a face for two cells $T_i\in\T_{i,h}$ and  $T_j\in\T_{j,h}$, $i\neq j$,
the diffusive and convective fluxes are assigned to $K$ from both sides of $\Gamma_h$.
The discrete net flux $F_{\Gamma_h}(u_h)$ at the barycenter of $K$ is computed as the jump of the fluxes over $K$.
In the TraceFEM this value is assigned to all $\bx\in K$, and numerical integration is  done over all surface elements $K\in\F_h$ to compute the right-hand side of the algebraic system.

To satisfy all (discretized) equations and boundary conditions we iterate between the bulk FV and surface FE solvers on each time step. %For the presentation we suppress further the subindex $h$, which indicates mesh objects, and
We assume an implicit time stepping method (in experiments we use backward Euler). This results in the following system on each time step.
\begin{equation} \label{system}
\left\{
\begin{aligned}
 \cL u:= \tilde\phi u + \Div(\bw  u- D \nabla) u &= \hat{f} \quad \text{in}~ \Omega\setminus\Gamma,\\
 u&=v \quad \text{on}~ \Gamma,\\
 {D}\bn_\dO\cdot\nabla u=0 \quad \text{on}~ \dO_N,\quad u&=u_D \quad \text{on}~ \dO_D,\\
 \cL_\Gamma v:=\tilde\phi_\Gamma v + \Div_\Gamma (\bw_\Gamma v - d D_\Gamma \nabla_\Gamma v) &=  F_\Gamma(u)+ \hat{f}_\Gamma  \quad \text{on}~~\Gamma,\\
 {D}_\Gamma\bn_\dG\cdot\nabla v=0 \quad \text{on}~ \dG_N,\quad v&=v_D \quad \text{on}~ \dG_D,
 \end{aligned}
\right.
\end{equation}
the right hand sides $\hat{f}$ and $\hat{f}_\Gamma$ account for the  solution values at the previous time step. Note that condition \eqref{cont_v} is satisfied by the construction of trace spaces in the finite element method
and condition \eqref{cont_F} is accounted weakly by the TraceFEM variational formulation.

{
We solve the coupled system \eqref{system} by the fixed point method: Given ${u}^0,{v}^0$, the initial guess,  we iterate for $k=0,1,2,\dots$ until convergence: \\
Step 1: Solve for ${u}^{k+1}$,
\begin{equation} \label{step1}
\left\{
\begin{aligned}
 \cL {u}^{k+1}&= \hat{f}~\text{in}~ \Omega\setminus\Gamma,\quad {u}^{k+1}={v}^{k} \quad \text{on}~ \Gamma,\\
 {D}\bn_\dO\cdot\nabla {u}^{k+1}&=0 \quad \text{on}~ \dO_N,\quad {u}^{k+1}=u_D \quad \text{on}~ \dO_D,\\
 \end{aligned}
\right.
\end{equation}
Step 2: Solve for ${v}^{\rm aux}$ and update for ${v}^{k+1}$ with a relaxation parameter $\omega$,
\begin{equation} \label{step2}
\left\{
\begin{aligned}
 \cL_\Gamma {v}^{\rm aux}&=  F_\Gamma(u^{k+1})+ \hat{f}_\Gamma  \quad \text{on}~~\Gamma,\\
 {D}_\Gamma\bn_\dG\cdot\nabla_\Gamma {v}^{\rm aux}&=0 \quad \text{on}~ \dG_N,\quad {v}^{\rm aux}=v_D \quad \text{on}~ \dG_D\\
 {v}^{k+1}&=\omega{v}^{\rm aux}+(1-\omega) {v}^{k},\quad \omega\in(0,1],
 \end{aligned}
\right.
\end{equation}
}

\begin{remark}\rm \label{rem1} {Below we show that the fixed point method is equivalent to a preconditioned Richardson iteration for the discrete Poincar\'{e}--Steklov operator. Assume that $\cL$ is linear (this is true for our differential model, but
the particular FV discretization applied here is actually non-linear).
Let's} split $u=u_0+\hat{u}$, $v=v_0+\hat{v}$, where $u_0,\,v_0$ satisfy
\[
\left\{
\begin{aligned}
\cL u_0&= \hat{f}~\text{in}~ \Omega\setminus\Gamma,\quad u_0=0 \quad \text{on}~ \Gamma,\\
 D\bn_\dO\cdot\nabla u_0&=0 \quad \text{on}~ \dO_N,\quad u_0=u_D \quad \text{on}~ \dO_D,
  \end{aligned}
\right.\qquad
\left\{
\begin{aligned}
 \cL_\Gamma {v}_0&=  0  \quad \text{on}~~\Gamma,\\
 D_\Gamma\bn_\dG\cdot\nabla {v}_0&=0 \quad \text{on}~ \dG_N,\quad {v}_0=v_D \quad \text{on}~ \dG_D
 \end{aligned}
\right.
\]
Now the iterations \eqref{step1}--\eqref{step2} can be written in terms of  $\hat{u}$ and $\hat{v}$ parts of the bulk and surface  concentrations:
\begin{equation} \label{step1a}
\begin{split}
\left\{
\begin{aligned}
 \cL \hat{u}^{k+1}&= 0~\text{in}~ \Omega\setminus\Gamma,\quad \hat{u}^{k+1}=\hat{v}^{k} \quad \text{on}~ \Gamma,\\
 {D}\bn_\dO\cdot\nabla \hat{u}^{k+1}&=0 \quad \text{on}~ \dO_N,\quad \hat{u}^{k+1}=0 \quad \text{on}~ \dO_D,\\
 \end{aligned}
\right. \\
\left\{
\begin{aligned}
 \cL_\Gamma \hat{v}^{\rm aux}&=  F_\Gamma(u^{k+1})+ \hat{f}_\Gamma  \quad \text{on}~~\Gamma,\\
 {D}_\Gamma\bn_\dG\cdot\nabla_\Gamma \hat{v}^{\rm aux}&=0 \quad \text{on}~ \dG_N,\quad \hat{v}^{\rm aux}=0 \quad \text{on}~ \dG_D\\
 \hat{v}^{k+1}&=\omega\hat{v}^{\rm aux}+(1-\omega) \hat{v}^{k},\quad \omega\in(0,1].
 \end{aligned}
\right.
\end{split}
\end{equation}
%{with $u^{k+1}=u_0+\hat{u}^{k+1}$.}
Now we note that $\hat{u}$ is a (generalized) harmonic extension of $\hat{v}$  on $\Omega\setminus\Gamma$ and $S_\Gamma\,:\,\hat{v}\to F_\Gamma(\hat{u})$ is the  Dirichlet to Neumann (discrete) Poincar\'{e}--Steklov operator. Using this notation,
one can write the surface equation for $\hat{v}$ in the compact operator form,
\begin{equation} \label{PS}
(\cL_{\Gamma,0}-S_\Gamma) \hat{v}= \hat{F}\quad \text{on}~~\Gamma,\quad\text{with}~~ \hat{F}:=F_\Gamma(u_0)+ \hat{f}_\Gamma.
\end{equation}
We use zero index in $\cL_{\Gamma,0}$ to stress that the operator accounts for homogenous boundary conditions on $\dG$. It is easy to see that \eqref{step1a} is the Richardson iterative process for the surface
equation \eqref{PS}, with the preconditioner $W=\cL_{\Gamma,0}^{-1}$ and the relaxation parameter $\omega$:
\begin{equation} \label{Richards}
\hat{v}^{k+1}=\hat{v}^{k}-\omega\,W\left(\, (\cL_\Gamma-S_\Gamma) \hat{v}^k-\hat{F}\right),\quad k=0,1,2,\dots.
\end{equation}
From \eqref{Richards} we see that a more efficient iterative process based on a different choice
of preconditioner and employing  Krylov subspaces may be feasible {(if $\cL$ is non-linear one may consider Anderson's mixing to accelerate convergence)}.  However, we do not pursue this topic further in this paper.
\end{remark}

%%%%%%%%%%%%%%%%%%%%%%%%%%%%%%%%%%%%%%%%%%%%%%%%%%%%%%%%%%%%%%%%%%%%%%%%%%%%%%%%%%%%%%%%%%%%%%%%%%%%%%%%%%%%%%%%%%%
%%%%%%%%%%%%%%%%%%%%%%%%%%%%%%%%%%%%%%%%%%%%%%%%%%%%%%%%%%%%%%%%%%%%%%%%%%%%%%%%%%%%%%%%%%%%%%%%%%%%%%%%%%%%%%%%%%%

\section{Numerical results and discussion}\label{s_numer}
This section collects several numerical examples, which demonstrate the accuracy and capability of the hybrid method.
 We perform a series of tests, where we simulate steady and time-dependent solutions  in a bulk domain with an imbedded  fracture. {We also include an a example with a smooth curved surface (a sphere) embedded in a bulk domain and a given analytical solution for a surface-bulk problem with Henry interface condition.} %common for surfactant transport in two-phase flows.
 {To measure the error we shall use $L^2$, $H^1$ and $L^\infty$ surface and volume norms.
%$L^\infty$-norm is defined in the obvious way as the maximum norm over concentration nodes in $\Omega$ and all quadrature points over $\Gamma$.
For the computed solutions $u_h,v_h$ and true solutions $u,v$, these norms are defined below. In the volume, we set
\[
\begin{split}
&\mbox{err}_{L^\infty(\Omega)}:=\max_{T\in\T_h}|u_h(\bx_T)-u(\bx_T)|,\quad \mbox{err}_{L^2(\Omega)}:=\left(\sum_{T\in\T_h}\text{vol}(T)|u_h(\bx_T)-u(\bx_T)|^2\right)^{\frac12},\\ &\mbox{err}_{H^1(\Omega)}:=\left(\sum_{T\in\T_h}\text{vol}(T)|\nabla I(u_h)(\bx_T)-\nabla u(\bx_T)|^2\right)^{\frac12},
\end{split}
\]
where $I(u_h)$ is the $P_1$ least-square interpolant to the values of $u_h$ in barycenters of the cells from $\omega(\bx_T)\cap\Omega_i$, for $T\in\Omega_i$. Over the surface, we set
\[
\mbox{err}_{L^\infty(\Gamma)}:=\max_{\Gamma}|v_h-v^e|,\quad \mbox{err}_{L^2(\Gamma)}:=\|v_h-v^e\|_{L^2(\Gamma)},\quad
\mbox{err}_{H^1(\Gamma)}:=\|\nabla_{\Gamma_h}v_h-\nabla_{\Gamma_h}v\|_{L^2(\Gamma)},
\]
where $v^e$ is the  extension of $v$ from $\Gamma$ to $\Gamma_h$ along normal directions to $\Gamma$.
}

\subsection{Steady analytical solution for a  triple fracture problem}\label{sec_ex2}

\begin{figure}[ht!]
\begin{center}
\includegraphics[width=0.31\textwidth]{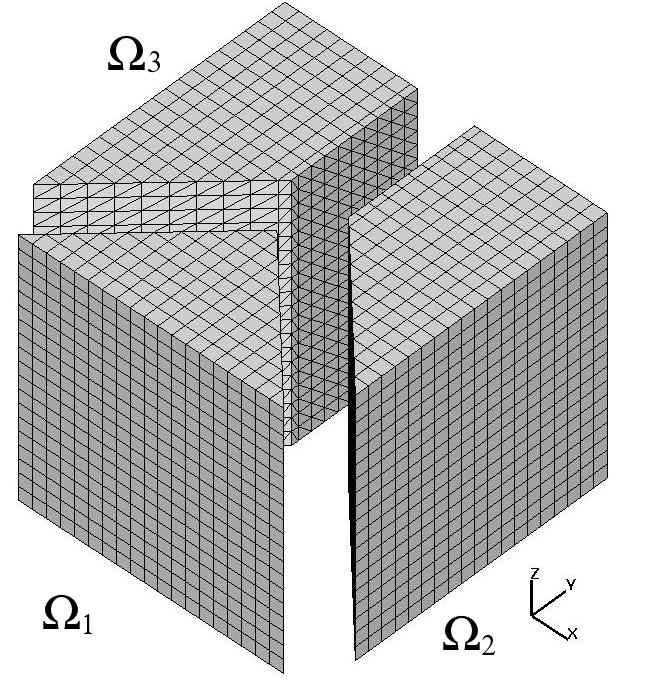} \includegraphics[width=0.36\textwidth]{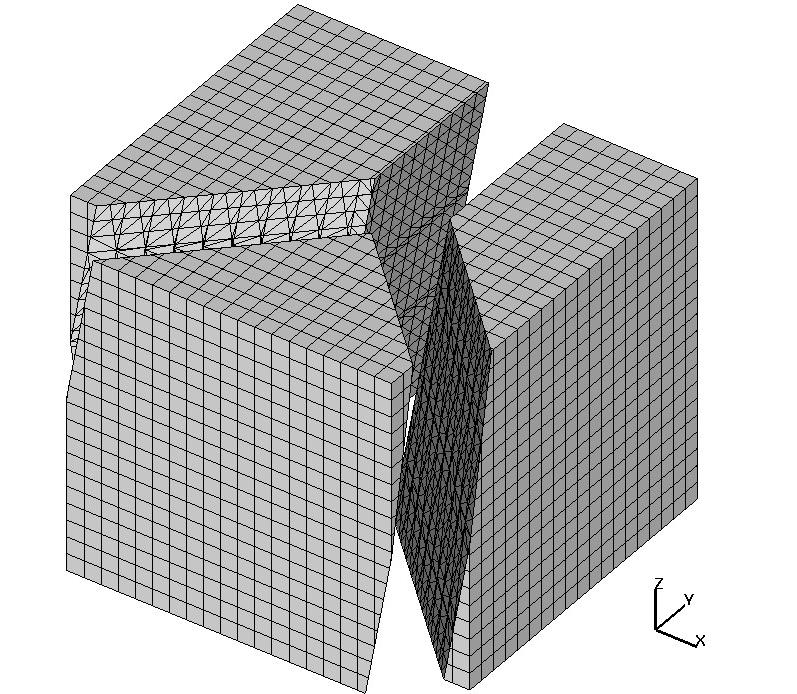} \includegraphics[width=0.31\textwidth]{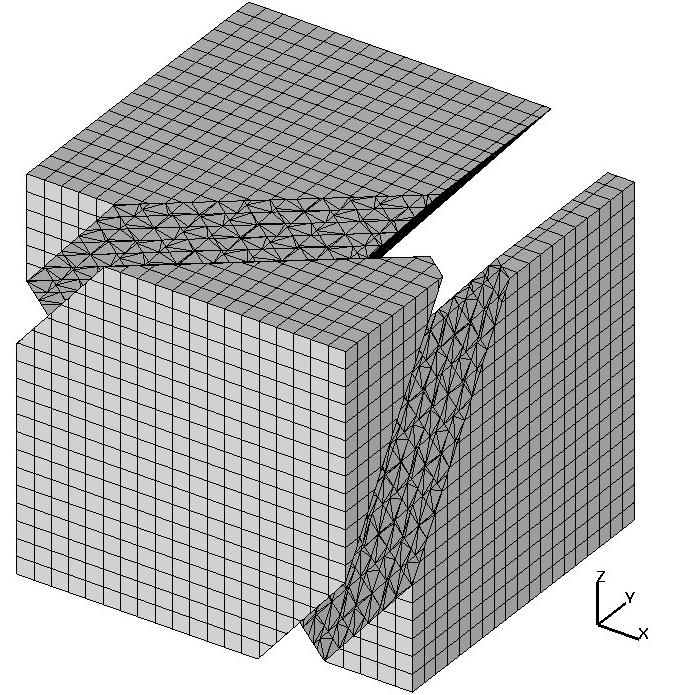}
\caption{\label{fig:triple}
The figure illustrates the bulk domain with uniform mesh and the fracture. On the left picture the fracture is set orthogonal to the $xy$-plane, while on the middle and right pictures the fracture is rotated by 20 and 40 degrees.}
\end{center}
\end{figure}

Our next experiment deals with the coupled surface--bulk diffusion problem in the domain $\Omega = [0,1]^3$ with
an embedded piecewise planar $\Gamma$. We design $\Gamma$ to model a branching fracture.  In the basic model,
$\Gamma=\Gamma(0)$ consists of three planar pieces,
\[\Gamma(0) = \Gamma_{12}\cup\Gamma_{13}\cup\Gamma_{23},\quad \Gamma_{ij}=\overline{\Omega_i}\cap\overline{\Omega_j}~~i\neq j,
\]
such that
\[
\Omega_{1}=\{\bx\in\Omega\,|\, x<\frac12~\text{and}~ y>x\},\quad \Omega_{2}=\{\bx\in\Omega\,|\, x>\frac12~\text{and}~y>x-1\},\quad
\Omega_{3}=\Omega\setminus(\overline{\Omega}_{1}\cup\overline{\Omega}_2).
\]
This subdivision is illustrated in Figure~\ref{fig:triple} (left). The pieces $\Gamma_{ij}$ belong to certain planes of symmetry
for the cube, and so the induced triangulation of $\Gamma(0)$ and the cut cells in the bulk domain are all quite regular.
To model a generic situation when $\Gamma$ cuts through the background mesh in an arbitrary way, we  consider other
tessellations of $\Omega = [0,1]^3$ into three subdomains by a surface $\Gamma(\alpha)$. The surface $\Gamma(\alpha)$ is obtained from $\Gamma(0)$ by applying the clockwise rotation by the angle $\alpha$ around the axis $x=z=0.5$. We take $\alpha=20^o$ and
$\alpha=40^o$, the resulting tessellations of $\Omega$ are illustrated in  Figure~\ref{fig:triple} (middle and right pictures).
More precisely, we define
\[
\Gamma(\alpha)=\{\bx\in\Omega\,|\, \by\in\Gamma(0),~\by-\bx_0=\mathcal{Q}_\alpha(\bx-\bx_0)\},~~\text{with}~
\mathcal{Q}_\alpha=\begin{bmatrix}\cos\alpha&0&-\sin\alpha\\ 0&1&0\\ \sin\alpha&0&\cos\alpha\end{bmatrix},~ \bx_0=({\footnotesize\frac12},0,{\footnotesize\frac12})^T.
\]

\begin{figure}[ht!]
\begin{center}
\includegraphics[width=0.32\textwidth]{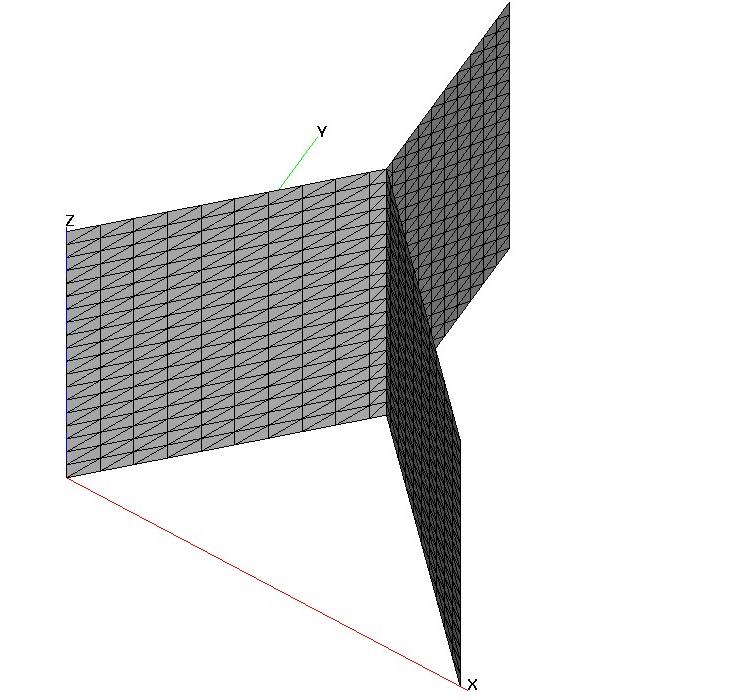}  \includegraphics[width=0.32\textwidth]{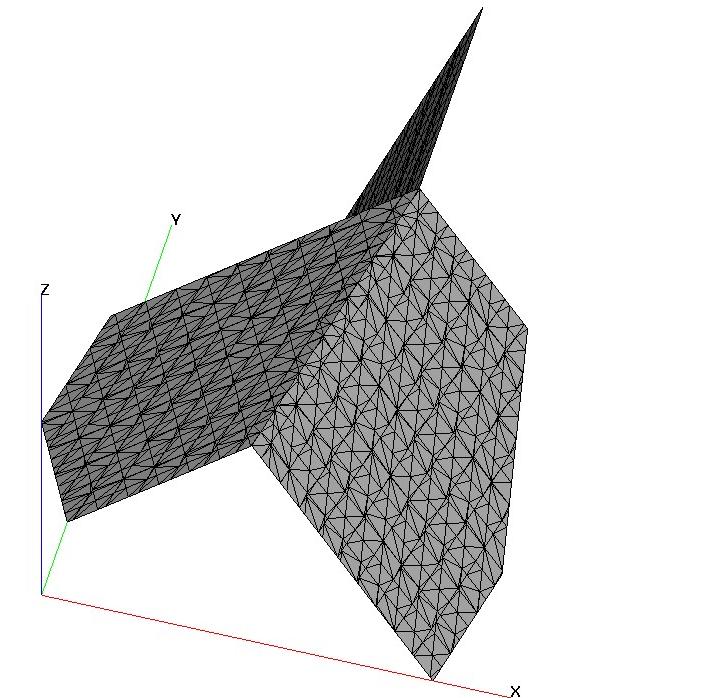}
\includegraphics[width=0.32\textwidth]{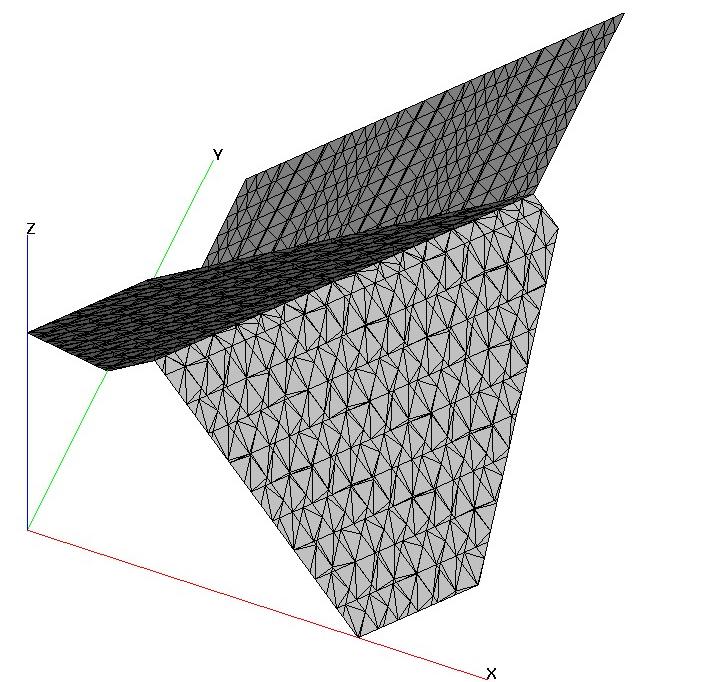}
\caption{\label{fig:frac}
The figure illustrates the induced surface mesh on the fracture, when it cuts through the uniform bulk mesh in different ways.}
\end{center}
\end{figure}

Similar to the series of numerical experiments with the embedded spherical $\Gamma$, here we set   the source terms $f_i$ and $f_\Gamma$ and the boundary conditions such that the solution to the stationary problem \eqref{diffeq1}--\eqref{bc} is known.  To define the  solution $\{v,u\}$ solving the stationary equations \eqref{diffeq1}--\eqref{bc}, we first introduce
\begin{equation}\nonumber
\psi_1 = \left\{
\begin{array}{lr}
 16(y-\frac12)^4,& y>\frac12\\
 0, & y \le \frac12\\
\end{array}
\right.,\quad\psi_2 = x -y,\quad \psi_3 = x + y - 1.
\end{equation}
We define the solution of the basic model problem ($\alpha=0$)
\begin{equation}\nonumber
u(\bx)=\left\{
 \begin{array}{rl r}
  & \sin(2\pi z)\cdot\psi_2(\bx) \cdot\phi_3(\bx) & \bx\in \Omega_1,\\
  & \sin(2\pi z)\cdot\psi_1(\bx) & \bx\in \Omega_2,\\
  & \sin(2\pi z)2x\cdot\psi_1(\bx) & \bx\in \Omega_3,
 \end{array}
 \right.\qquad v=u|_{\Gamma(0)}.
\end{equation}
Note that the constructed exact solution is continuous across $\Gamma(0)$, but the normal derivatives are discontinuous.
Other parameters in \eqref{diffeq1}--\eqref{diffeq2} are set to be  $\mathbf{w}=\mathbf{w}_\Gamma=0$, $\phi_1=\phi_2=\phi_\Gamma=0$, $D_1=D_2=D_\Gamma=I$, and $d=1$. For the problem setup with the rotated fracture, $\alpha>0$ we set the exact solution
$v_\alpha(\bx)=v(\by)$, $u_\alpha(\bx)=u(\by) $, with $\by=\mathcal{Q}_\alpha(\bx-({\small\frac12},0,{\small\frac12})^T)$.

\begin{table}[t]
\begin{center}
\caption{The error in the numerical solution for the steady problem with  triple fracture, $\alpha = 0$.
\label{tab:triple0}}\smallskip
\small
\begin{tabular}{rr|llllll}\hline
&\#d.o.f. & $L^2$-norm & rate & $H^1$-norm& rate & $L^\infty$-norm& rate \\ \hline\\[-2ex]
{$\Omega$}
&855&	6.374e-3     &        &4.214e-1&          & 3.920e-2&      \\
&7410&	 1.698e-3    & 1.84   &1.631e-1&1.36      & 1.276e-2& 1.56 \\
&61620&	4.235e-4     & 1.97   &6.193e-2&1.39      & 3.506e-3& 1.83 \\
&502440& 1.044e-4    & 2.00   &2.348e-2&1.40      & 1.129e-3& 1.62 \\ \hline\\[-2ex]
{$\Gamma$}
&232& 8.469e-3       &         &2.914e-1 &          &9.280e-3&     \\
&1242&2.003e-3	       & 1.79  &1.387e-1 & 0.92     &2.779e-3& 1.44\\
&5662&5.588e-4	       & 1.84  &6.874e-2 & 1.01     &1.217e-3& 1.09\\
&24102&1.791e-4       &  1.64  &3.395e-2 & 1.02     &5.181e-4& 1.18\\  \hline
\end{tabular}
\end{center}
\end{table}

\begin{table}[t]
\begin{center}
\caption{The error in the numerical solution for the steady problem with  triple fracture, $\alpha = 20$.
\label{tab:triple20}}\smallskip
\small
\begin{tabular}{rr|llllll}\hline
&\#d.o.f. & $L^2$-norm & rate & $H^1$-norm& rate & $L^\infty$-norm& rate \\ \hline\\[-2ex]
{$\Omega$}
&965&6.319e-3&               &4.208e-1&        &3.754e-2  &    \\
&7872&1.805e-3      & 1.79   &1.661e-1& 1.34   &1.280e-2  &1.55\\
&63592& 5.623e-4    & 1.80   &6.371e-2& 1.38   & 3.411e-3 &1.90\\
&510390&1.602e-4    & 1.81   &2.442e-2& 1.39   & 1.146e-3 &1.57\\\hline\\[-2ex]
{$\Gamma$}
&321& 7.792e-3    &       &2.694e-1&        &2.716e-2&     \\
&1692&2.084e-3    & 1.59  &1.240e-1& 1.12   &5.400e-3& 1.94\\
&7944&7.019e-4    & 1.41  &6.291e-2& 0.98   &2.001e-3& 1.29\\
&{33272}&{2.441e-4}   & {1.52}  &{3.173e-2}& {0.99}   &{7.217e-4}& {1.47}\\  \hline
\end{tabular}
\end{center}
\end{table}

\begin{table}[t]
\begin{center}
\caption{The error in the numerical solution for the steady problem with  triple fracture, $\alpha = 40$.
\label{tab:triple40}}\smallskip
\small
\begin{tabular}{rr|llllll}\hline
&\#d.o.f. & $L^2$-norm & rate & $H^1$-norm& rate & $L^\infty$-norm& rate \\ \hline\\[-2ex]
{$\Omega$}
&991&	 5.934e-3   &       & 4.080e-1&      &  3.783e-2 &     \\
&7996&	 1.700e-3   & 1.80  & 1.621e-1& 1.33 &  1.276e-2 & 1.56\\
&64046&  4.907e-4   & 1.80  & 6.263e-2& 1.37 &  3.515e-3 & 1.86\\
&512258& 1.503e-4   & 1.82  & 2.541e-2& 1.39 &  1.237e-3 & 1.61\\  \hline\\[-2ex]
{$\Gamma$}
&353&     8.167e-3 &       &2.709e-1&     & 2.696e-2 &      \\
&1932&	  2.146e-3 & 1.66  &1.275e-1& 1.09& 5.566e-3 & 1.85 \\
&8766&    7.115e-4 & 1.59  &6.279e-2& 1.02& 2.063e-3 & 1.31 \\
&{36676}&   {2.538e-4} & {1.49}  &{3.121e-2}& {1.01}& {7.251e-4} & {1.51} \\  \hline
\end{tabular}
\end{center}
\end{table}
\label{s_ex2}

The numerical results for this coupled problem with the triple fracture problem are reported in Tables~\ref{tab:triple0}--\ref{tab:triple40}.  We observe stable convergent results for $\alpha=0$ as well as for more general case of $\alpha>0$.   An interesting feature of this problem is that the surface $\Gamma$ is
only piecewise smooth.  The bulk grid is not fitted to the internal edge $\mathcal{E}=\Gamma_{12}\cap\Gamma_{13}\cap\Gamma_{23}$,
and hence the tangential derivatives of $v$ are discontinuous inside certain cubic cells from $\T_h^\Gamma$.
{Therefore, a kink in $v$ cannot be represented by the finite element approximation.} This {may} result in  {a}  reduction of convergence order. {Both the performance of the FV method for cut  cells (cut cells inherit a regular structure from the background mesh for $\alpha=0$, but are very irregular for $\alpha>0$ ) and the presence of the kink influences the observed convergence rates. }

\begin{table}[h!]   {
\begin{center}
\caption{Iteration numbers in \eqref{step1}--\eqref{step2} for the steady problem example in section~\ref{sec_ex2}.
\label{t_conv}}\smallskip
\small
\begin{tabular}{r|lll}\hline
ref. level. & $\alpha=0$ & $\alpha=20^o$ & $\alpha=40^o$ \\ \hline\\[-2ex]
0&	22   & 74  &24    \\
1&	29   & 90  &32    \\
2&	212  & 325 &228    \\
3& 782   & 917 &851   \\ \hline
\end{tabular}
\end{center}    }
\end{table}

{
Finally, Table~\ref{t_conv} shows the performance of the fixed-point iteration~\eqref{step1}--\eqref{step2}. We set $\omega=1$ and take $u^0=0$, $v^0=0$. \rev{The solver is stopped after a relative reduction of the Euclidean}  norm of  both surface and bulk equations residuals by a factor of $10^4$ (a stronger convergence criterion was not found to improve solution accuracy). \rev{In} each outer iteration, the surface linear subproblem was solved by exact factorization, while a few Picard iterations with exact factorization of linearized problem were done to solve the bulk system in \eqref{step1}.   The solver does not scale in an optimal way with respect to the mesh size and more research is needed to improve its performance, cf. Remark~\ref{rem1}. We postpone this topic for the future research. We also note that for time dependent problems studied  below including time-dependent terms and taking initial guess to be the solution from the previous time step improves convergence of \eqref{step1}--\eqref{step2} a lot,
and we typically need 1 or 2 iterations \rev{for} each time step.}

\subsection{Propagating front in the porous medium with triple fracture}
\definecolor{linkcolor}{HTML}{799B03} % цвет ссылок
\definecolor{urlcolor}{HTML}{799B03} % цвет гиперссылок

\hypersetup{pdfstartview=FitH, linkcolor=linkcolor,urlcolor=urlcolor, colorlinks=true}

\begin{figure}[h!]
\begin{center}
a)~~\includegraphics[width=0.4\textwidth]{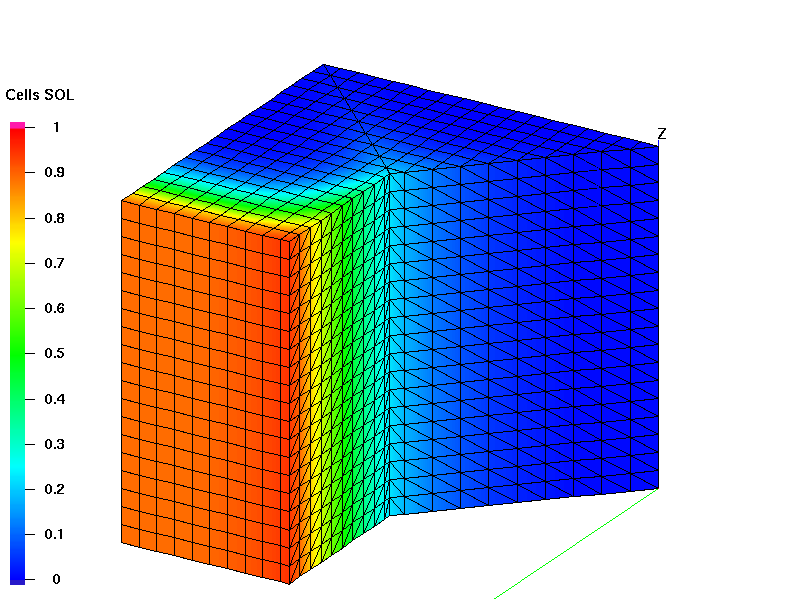}\qquad b)~~\includegraphics[width=0.4\textwidth]{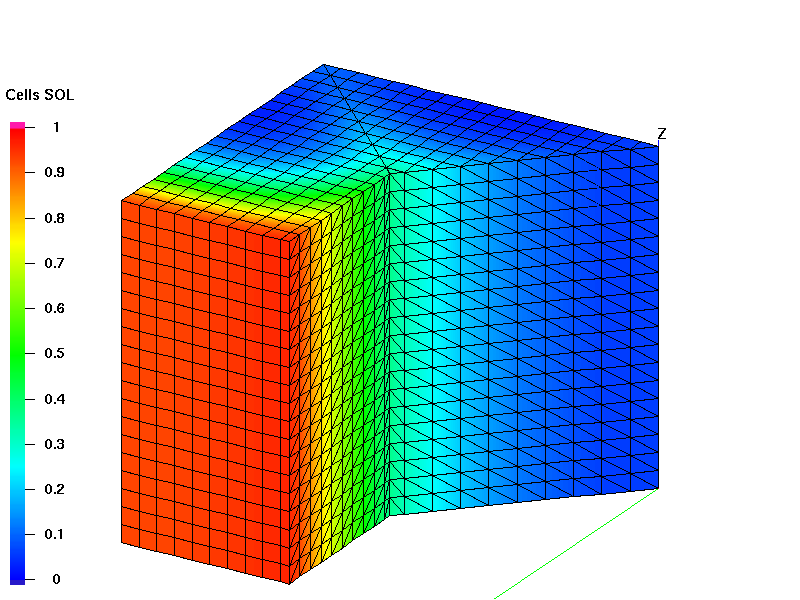}\\
c)~~\href{www.math.uh.edu/~molshan/OLSH/pdf_refs/pdf_ref_video_outx4.html}{\includegraphics[width=0.4\textwidth]{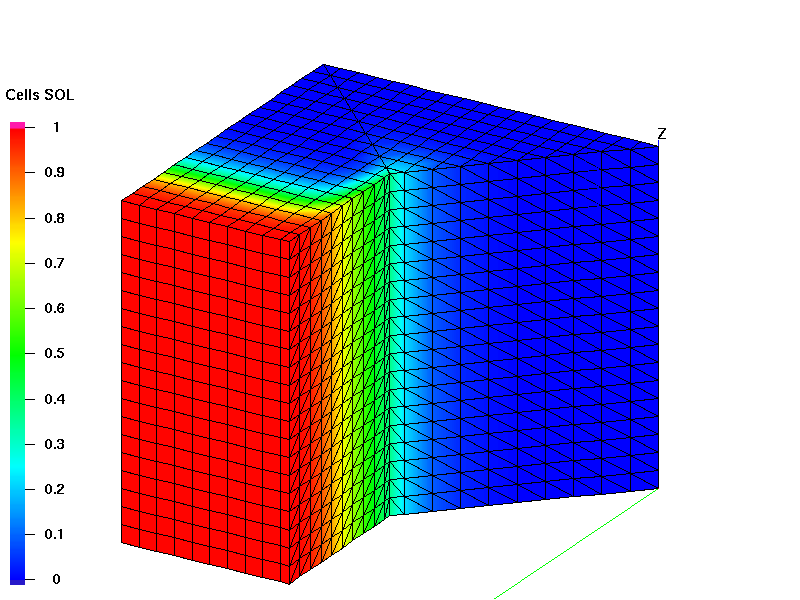}}\qquad d)~~\includegraphics[width=0.4\textwidth]{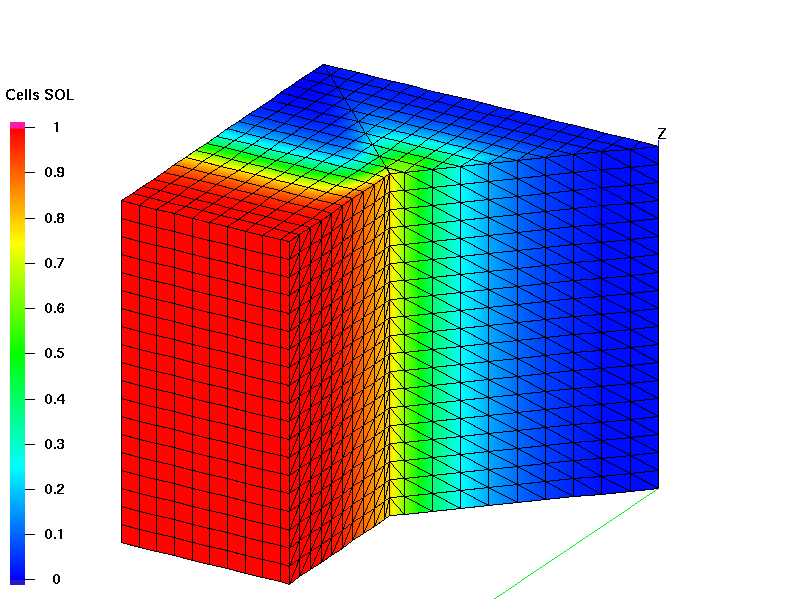}\\
\caption{\label{fig:prop}
The figures illustrates the propagating front of the concentration in the fracture and in the bulk: Pictures a), b) show dominating diffusion case, while  c), d) show dominating convection case. Pictures a) and c) {show} snapshots {of} the computed solution at time t=0.018, while pictures b) and d) snapshots the computed  solution at time t={0.033}. Click on picture~c) to run full animation of the experiment.}
\end{center}
\end{figure}

\begin{figure}[h!]
\begin{center}
a)~~\href{www.math.uh.edu/~molshan/OLSH/pdf_refs/pdf_ref_video_out_new_higher.html}{\includegraphics[width=0.4\textwidth]{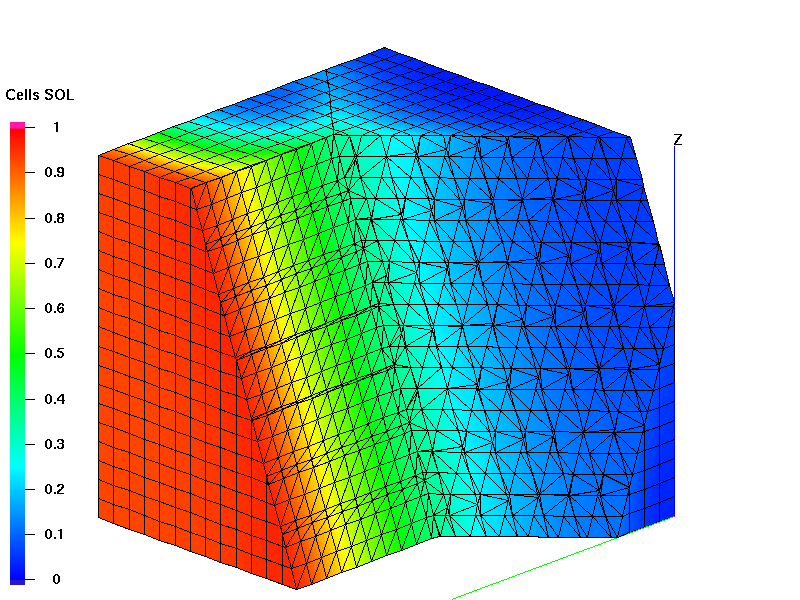}}\qquad b)~~\includegraphics[width=0.4\textwidth]{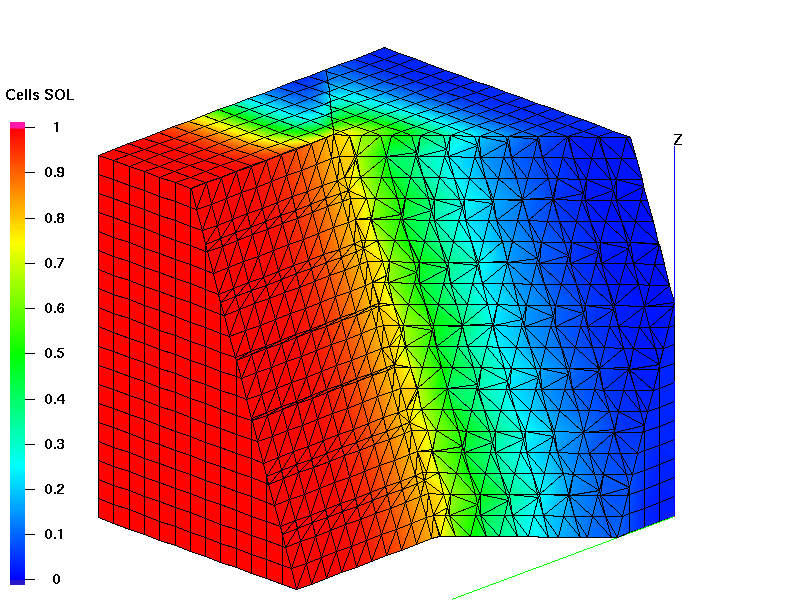}
\caption{\label{fig:prop1}
The figures illustrates the propagating front of the concentration in the fracture and in the bulk, with $\alpha=20^o$: Picture a) shows dominating diffusion case, picture b) shows dominating convection case;  both at time t=0.033. Click on picture~a) to run full animation of the experiment.}
\end{center}
\end{figure}

In the last series of experiments we compute the time dependent solution of \eqref{diffeq1}--\eqref{bc}. The bulk domain $\Omega$
and the fracture $\Gamma$ are the same as in the previous experiment in section~\ref{s_ex2}.
At  time
$t_0=0$ we set $u(t_0)=0$ in $\Omega$ and  $v(t_0)=0$ on $\Gamma$. On the face $\{y=1\}$ of the cube  we prescribe the constant concentration of a contaminant, while on other parts on the boundary the  diffusion flux is  set equal zero. Thus in \eqref{bc}, we have
\[
\begin{split}
\partial\Omega_D&=\partial\Omega\cap\{y=1\},\quad\partial\Omega_N=\partial\Omega\setminus\partial\Omega_D,\quad \partial\Gamma_D=\partial\Gamma\cap\{y=1\},\quad \partial\Gamma_N=\partial\Gamma\setminus\partial\Omega_D,\\  u_D&=1,\quad v_D=1,\quad u_0=0,\quad \text{and}\quad v_0=0.
\end{split}
\]
The time independent velocity field transports the contaminant in the bulk and along the fractures.
We set
\[
\begin{split}
\bw_i&=2\kappa(0,-1,0)^T,~~i=1,2,3,\quad\text{in}~\Omega\\
\bw_\Gamma&=5\kappa(0,-1,0)^T~~\quad\text{in}~\Gamma_{12},\quad
\bw_\Gamma=\kappa\mathcal{Q}_\alpha({\footnotesize\frac{1}{\sqrt{2}}},-{\footnotesize\frac{1}{\sqrt{2}}},0)^T
~~\quad\text{in}~\Gamma_{23},\quad
\bw_\Gamma=\kappa\mathcal{Q}_\alpha(-{\footnotesize\frac{1}{\sqrt{2}}},-{\footnotesize\frac{1}{\sqrt{2}}},0)^T~~\quad\text{in}~\Gamma_{13},
\end{split}
\]
where $\kappa\ge0$ is a parameter. One easily verifies the condition \eqref{cont_w} on the edge $\mathcal{E}=\Gamma_{12}\cap\Gamma_{13}\cap\Gamma_{23}$. Other parameters are set to be
\[
D_1=D_2=0.1\,I,\quad d=1,\quad D_\Gamma=I, \quad \phi_1=\phi_2=\phi_\Gamma=1.
\]

The computed solutions for $\kappa=1/8$ (diffusion dominated case) and $\kappa=8$ (convection plays a significant role)  are illustrated in Figures~\ref{fig:prop}--\ref{fig:prop1}. The fracture angle parameter $\alpha$ was set to $0$ and $20$ degrees, {respectively}.  %The full animation of the time dependent solutions can be found at \cite{TraceFEMurl}.

For this problem, the exact solution is not known. The computed solution occurs to be physically reasonable. We see no sign of spurious oscillations. {As expected, the contaminant propagates faster along the fractures.}

\subsection{Contaminant transport along the fracture.}

In this test we place  a continuous contaminant source on the upstream boundary of the fracture.
\begin{wrapfigure}{r}{0.45\textwidth}
\vspace{-20pt}
  \begin{center}
    \includegraphics[width=0.4\textwidth]{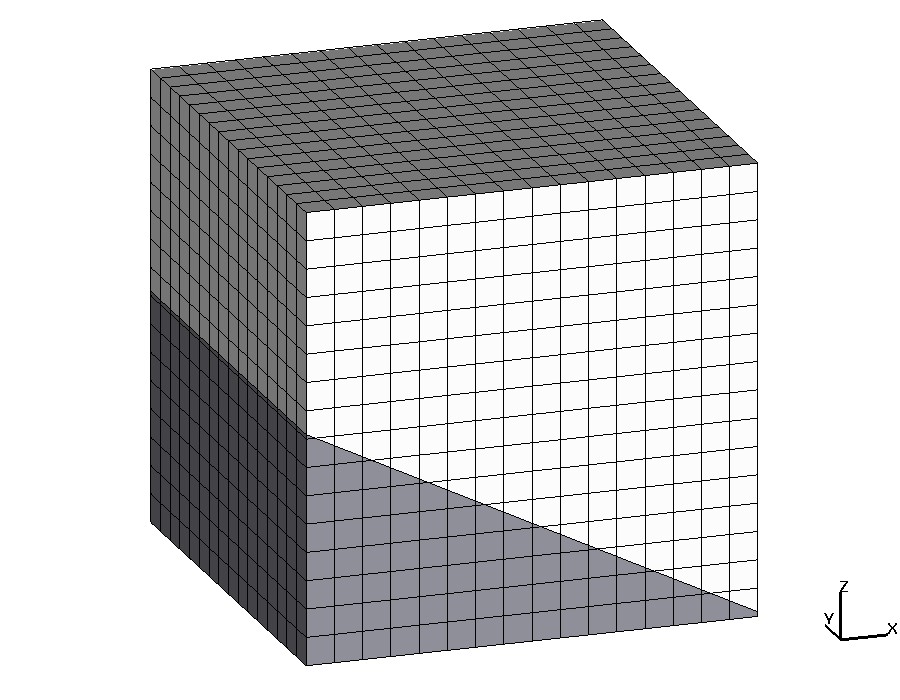}%{./Pictures/mainscreen1.png}
    \caption{The matrix and fracture in the test with contaminant transport along the fracture.}
    \label{fig:Front2}
  \end{center}
  \vspace{-20pt}
  \vspace{1pt}
\end{wrapfigure}
The matrix--fracture configuration for this test is shown in Figure~\ref{fig:Front2}, $\Omega=(0,1)^3$, and $\Gamma=\{\bx\in\Omega\,:\,z+\frac12x=0.51\}$. The boundary $x=0$
is inflow, in the fracture the wind is constant $\bw_\Gamma=(w_1,0,w_3)$, $|\bw_\Gamma|=1$, and the contaminant source occupies  the part of $\partial\Gamma$, {$\dG_D=\{(0,y,0.51)\,:\,y\in(\frac14,\frac34)\}$, $v_D=1$ on $\dG_D$.}
We assume that the  porous matrix  is almost impermeable and so we set $\bw_i=0$ in $\Omega_i$ (no flow in the rock) and $D_i=10^{-6}I$, $i=1,2$, {$\dO=\dO_N$}. In the fracture we assume isotropic diffusion with $D_\Gamma=10^{-4}I$. Other parameters are the same as in the previous test, {and $v=0$, $u=0$ at $t=0$}. Therefore, we expect that the contaminant transport happens along the fracture with very small diffusion to the porous matrix. This is a bulk--surface variant of a standard test case of numerical solvers for convection--diffusion problems~\cite{sun2013mathematical},  and one is typically interested in the ability of a method to capture the right position and the shape of the sharp propagating front and avoid spurious oscillations. For a comparison purpose, one may consider the exact solution for the problems posed
in a half-plane (or half-space) from~\cite{leij1990analytical,leij1991analytical}. This solution $C(x,y,0)$ is given in~\eqref{kapanalsol1}, it solves $C_t-D\Delta C+{C}_{x}=0$ in  $\widetilde \Omega = \{{(x,y)}\in\mathbb{R}^2\,:\, x>0\}$, with the
boundary condition
%\begin{equation*}
$
C(0,y,t)=\left\{
\begin{array}{c}
c_0, \textrm{  when } |y|<a,\\
0,  \textrm{  when } |y|>a,
\end{array}
\right.
$%\end{equation*}
and  initial conditions: $C(x,y,0)= 0$ in $\widetilde\Omega$.
\begin{equation}
 C(x,y,t)=\frac{x c_0}{(16 \pi D)^{\frac 12}} \int \limits_0^t  \tau^{-\frac 32}
\left\{
\mathbf{erf} \left[ \frac{a+y}{(4 D \tau)^\frac 12} \right]
+\mathbf{erf} \left[ \frac{a-y}{(4 D \tau)^\frac 12} \right]
\right\} \cdot \mathbf{exp} \left[ - \left( \frac{x-\tau}{(4D \tau)^{\frac 12}} \right) ^2 \right] d\tau.%
\label{kapanalsol1}
\end{equation}
where
\begin{equation*}%\label{errorfunc}
\mathbf{erf}(x) = \frac{2}{\sqrt{\pi}} \int \limits_0^x e^{-t^2} dt, \quad \mathbf{erfc}(x ) = 1-\mathbf{erf}(x)= \frac{2}{\sqrt{\pi}} \int \limits_x^\infty e^{-t^2} dt.
\end{equation*}

\begin{figure}[ht!]
\begin{center}
\includegraphics[width=0.8\textwidth]{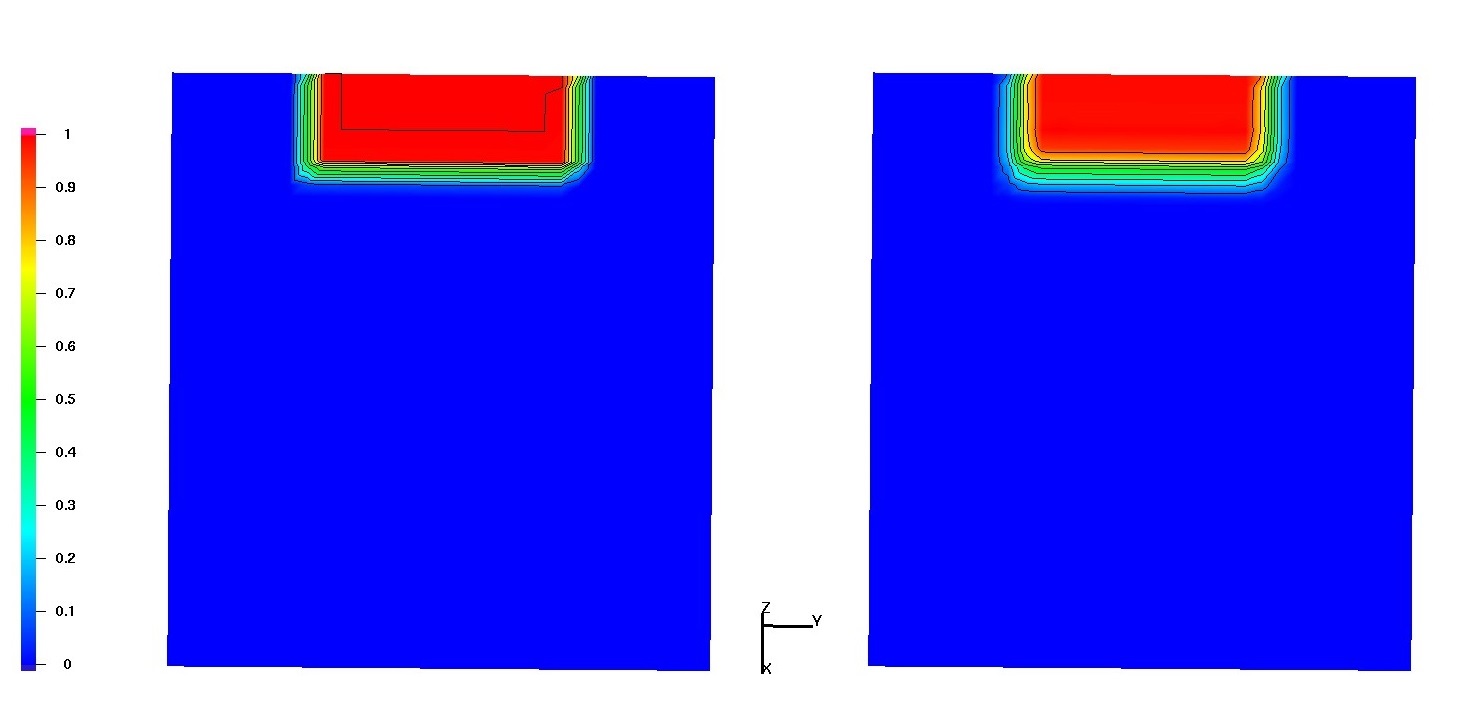}  \\
\includegraphics[width=0.8\textwidth]{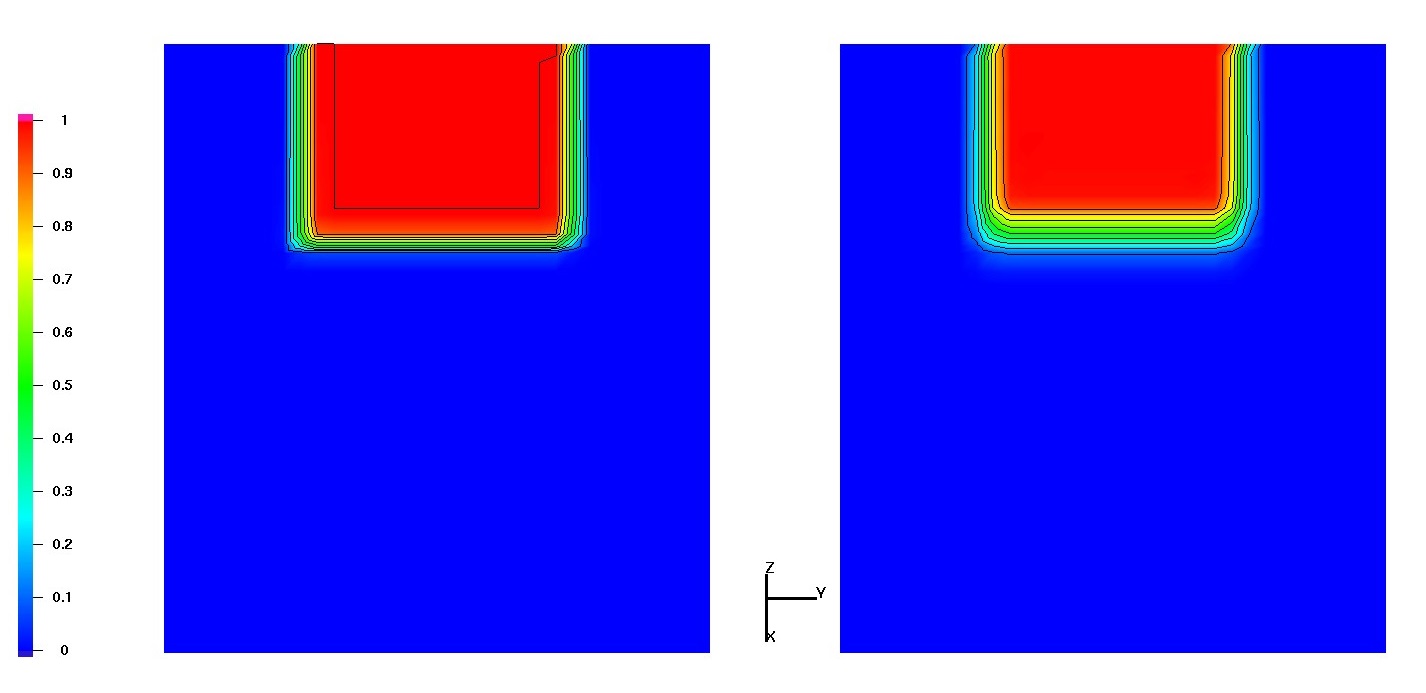}  \\
\includegraphics[width=0.8\textwidth]{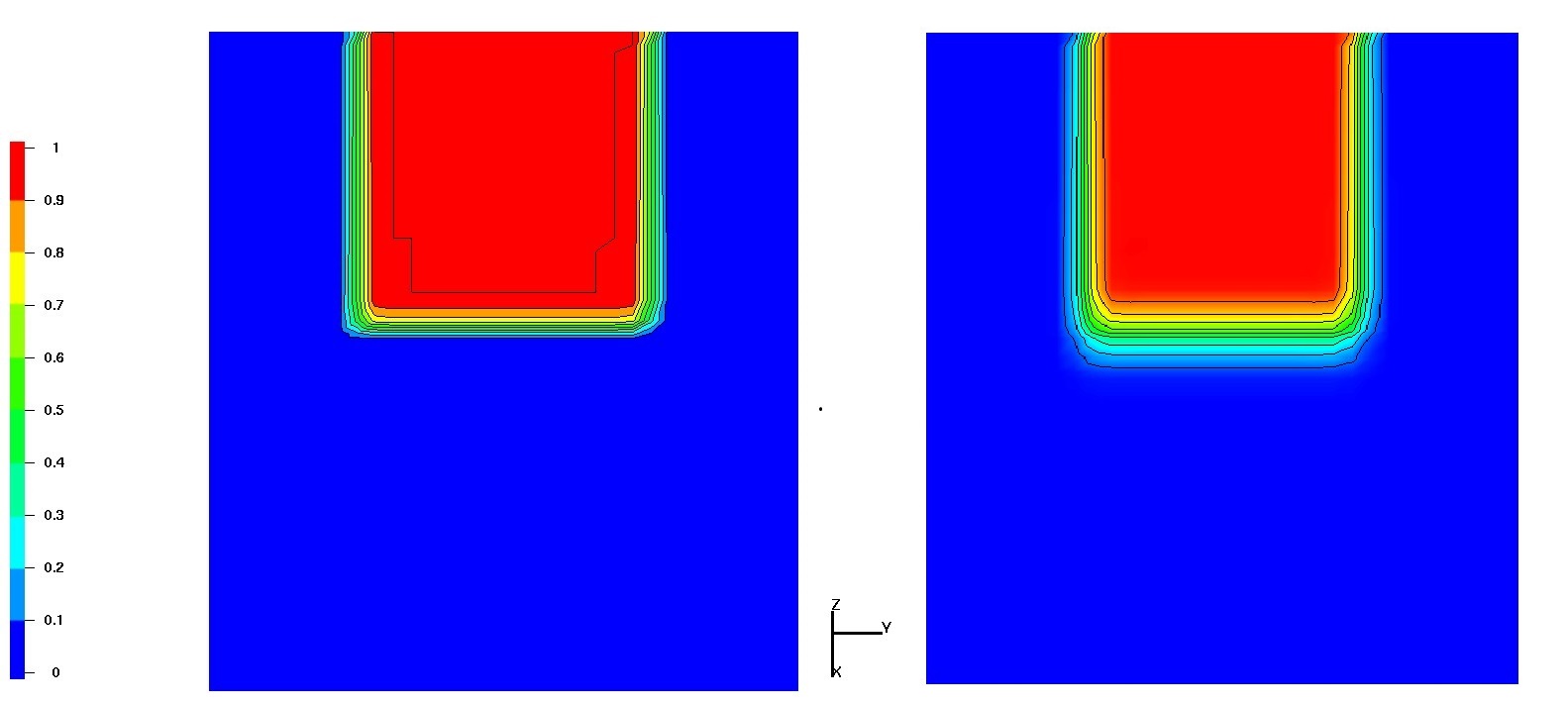}
\caption{\label{fig:rot20_3d}
Reference 2D solution  (left) and the fracture component of the computed  solution (right) for the contaminant transport along the fracture test
case. The solutions is shown for the times $t=0.17,\,0.34,\,0.5$.}
\end{center}
\end{figure}

We run our simulations with the uniform background mesh,  $h=\frac1{32}$, $\Delta t=10^{-2}$. The fracture cuts through the background mesh as illustrated in  Figure~\ref{fig:Front2} (for better visualization, this figure shows the background mesh for $h=\frac1{16}$). The computed solution and the `reference' solution is shown in  Figure~\ref{fig:rot20_3d} at several time instances. We recall that the coupled problem was solved and the contaminant also diffuses into the bulk, but this bulk diffusion was minor. We observe that the
computed solution well approximates the reference one; the computed front has the correct position and {is not  smeared too much. Moreover,} we do not observe  overshoots or undershoots in $v_h$.

{
\subsection{An example with a spherical drop immersed in a bulk}
We include one more test case but now with a different interface condition. This is
the instantaneous absorbtion--desorption condition \eqref{eq4a} with the Henry law to define $g_i$ and $f_i$.
This condition is common in the literature to model dissolvable surfactant transport in two-phase flows.
In this test from~\cite{gross2015trace} we consider a prototypical configuration for such models consisting of a spherical drop embedded in a cubic domain. We take $\Gamma$ to be the unit sphere centered at the origin and $\Omega = [-1.2,1.2]^3$.
By $\Omega_1$ we denote the interior of $\Gamma$, so $\Omega_1$ is the unit ball,  {$\Omega_2=\Omega\setminus\overline{\Omega}_1$}. For the velocity field we take a rotating field in the $x$-$z$ plane: $\bw=\frac{1}{10}(z,0,-x)$. This $\bw$ satisfies   $\Div\bw=0$ in $\Omega$ and $\bw \cdot \bn=0$ on $\Gamma$, i.e. the velocity field is everywhere tangential to the boundary and hence the steady interface is consistent with the kinematic condition: $\bw \cdot \bn$ is equal to the normal velocity of $\Gamma$ for immersible two-pase fluids, e.g.~\cite{GrossReuskenBook}. We set $\bw_i=\bw|_{\Omega_i}$ and $\bw_\Gamma=\bw|_{\Omega_\Gamma}$.

\begin{figure}[ht!]
\begin{center}
\includegraphics[width=0.45\textwidth]{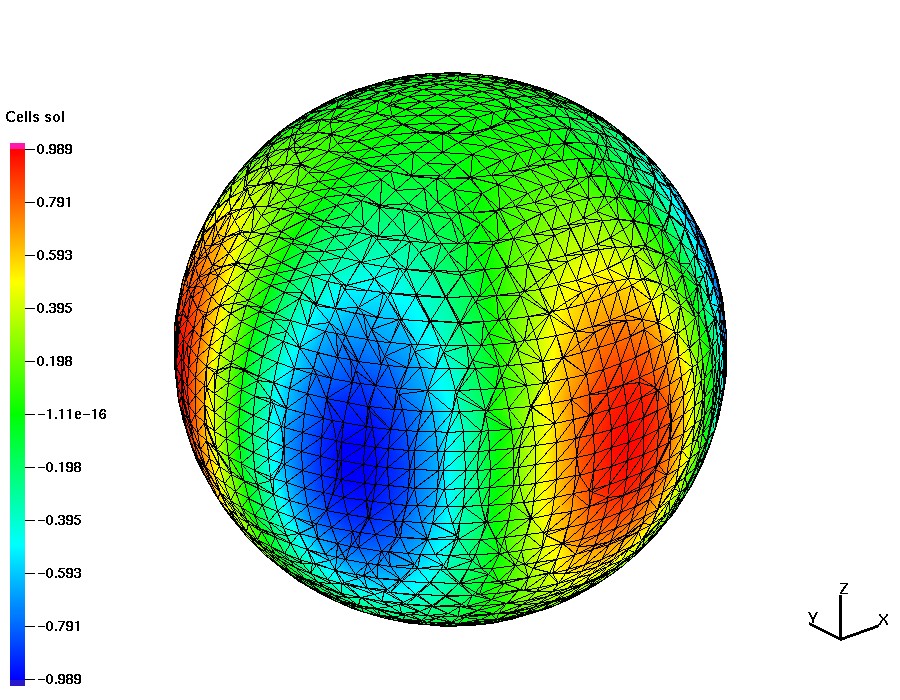}\qquad  \includegraphics[width=0.4\textwidth]{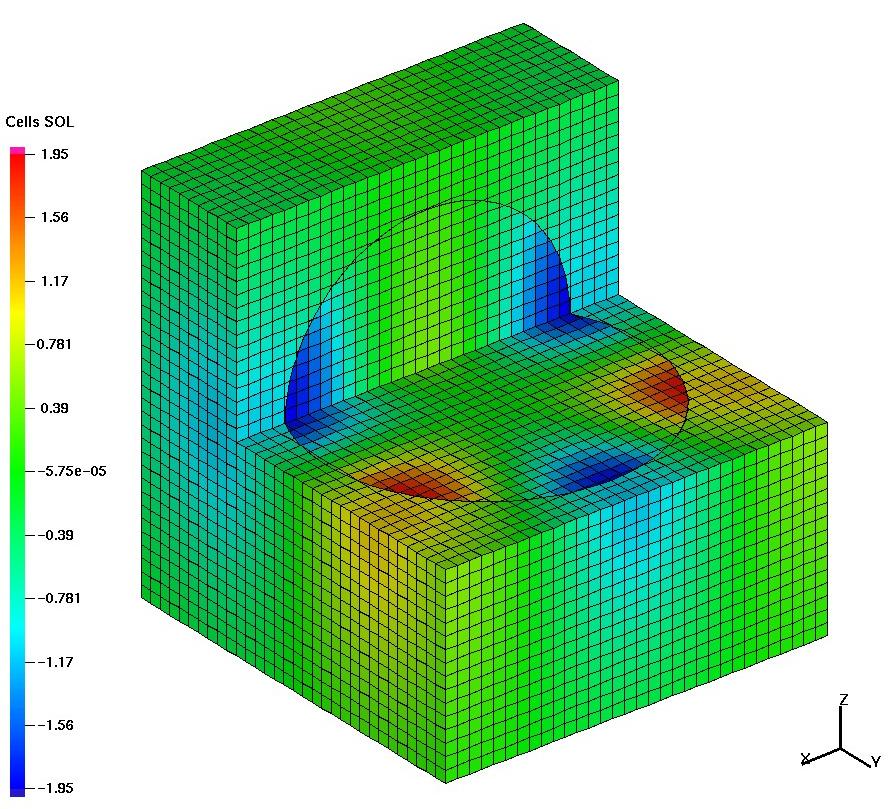}
\caption{\label{fig:sphere_loc}
Left: Induced surface mesh and the surface component of computed solution.
Right: Cut of the bulk mesh and the volume component of computed solution.}
\end{center}
\end{figure}

The material parameters are chosen as $D_1=0.5$, $D_2=1$, $D_\Gamma=1$ and $k_{1,a}=0.5$, $k_{2,a}=2$, $k_{1,d}=2$, $k_{2,d}=1$, $d=1$.
The source terms $f_i\in L^2(\Omega)$, $i=1,2$, and $f_\Gamma\in L^2(\Gamma)$   and  data on $\dO$ are taken such  that the exact solution of the  \textit{stationary} equations \eqref{diffeq1}--\eqref{diffeq2} is given by
\begin{equation*} %\label{exsol}
  %\begin{split}
  v(x,y,z) = 3x^2y - y^3, \quad
  u_1(x,y,z) = 2 u_2(x,y,z), \quad
  u_2(x,y,z) = e^{1-x^2-y^2-z^2} v(x,y,z).
%\end{split}
\end{equation*}
Since we solve for a steady-state solution, so we  set $\phi_1=\phi_2=\phi_\Gamma=0$.
 We prescribe Dirichlet boundary conditions on $\partial\Omega$, i.e. $\partial\Omega_N=\emptyset$,   $\partial\Gamma_N=\emptyset$, and $\partial\Gamma_D=\emptyset$ in \eqref{bc}. Conditions \eqref{cont_w}--\eqref{cont_F} for this test case are not relevant, since the surface is globally smooth and has no boundary.

\begin{table}
\begin{center}
\caption{Convergence of numerical solutions in the experiment with a spherical $\Gamma$ embedded in a cube.
\label{tab:sphere_uni}  }\smallskip
\small
\begin{tabular}{rr|llllll}\hline
&\#d.o.f. & $L^2$-norm & rate & $H^1$-norm& rate & $L^\infty$-norm& rate \\ \hline\\[-2ex]
{$\Omega$}
&736&	 3.223e-02 &        & 9.706e-01  &        &  1.072e-01  &      \\
&4920&	 6.687e-03 &    2.27& 1.555e-01  &    2.64&  3.799e-02  &1.50\\
&36088&  2.005e-03 &    1.74& 5.363e-02  &    1.55&  9.180e-01  &-4.59 \\
&275544& 5.055e-04 &    1.99& 1.825e-02  &    1.56&  2.777e-03  &8.37 \\  \hline\\[-2ex]

{$\Gamma$}
&460&      1.670e-02   &    &  2.065e-01&     & 3.863e-02&     \\
&1660&     4.037e-03   &2.05&  9.647e-02& 1.10& 1.060e-02&1.87 \\
&6628&     9.211e-04   &2.13&  4.745e-02& 1.02& 3.881e-03&1.45  \\
&26740&    2.457e-04   &1.91&  2.396e-02& 0.99& 8.875e-04&2.13  \\  \hline
\end{tabular}
\end{center}
\end{table}

In this set of experiments we take the sequence of uniform cubic meshes in $\Omega$, starting with $h=0.3$. The surface $\Gamma_h$ is reconstructed as described in section~\ref{s_rec} for $\phi(\bx)=1-|\bx|^2$. The computed solution as well as volume and induced surface meshes are illustrated in Figure~\ref{fig:sphere_loc}.
The computed errors for the bulk and surface concentrations are shown in Table~\ref{tab:sphere_uni}. For this example, the method
demonstrates  optimal convergence: $O(h)$ in the $H^1$ and $O(h^2)$ in the $L^2$ and surface norms. This is consistent with what is known about the convergence of the TraceFEM for linear bulk elements, see e.g. analysis and convergence rates for the same
experiment in \cite{gross2015trace}, where the TraceFEM has been used to discretized equations both on the surface and in the bulk.  For the volume component of the solution, the convergence is close to the second order in the $L^2$ norm and $1.5$ order in the $H^1$ norm.  It is also consistent with the results in \cite{Lipnikov:12}, where  super-convergence of the method in the $H^1$ norm was observed.  Convergence in the $L^\infty$-norm is somewhat less regular. We note that the $L^\infty$ convergence of the TraceFEM and of the non-linear FV method that we used has not been studied before.
\medskip

The aim of the next (final) test is to illustrate the performance of the method for the case of locally refined grids.  The setup is similar to the test with the sphere above, but the coefficients and the known solution are taken different to represent the situation of \rev{a} convection dominated problem with an internal layer. More precisely,  for the velocity field we take  $\bw=(-y\sqrt{1-z^2},x\sqrt{1-z^2},0)$, and set $\bw_i=\bw|_{\Omega_i}$ and $\bw_\Gamma=\bw|_{\Omega_\Gamma}$.

\begin{figure}[ht!]
\begin{center}
\includegraphics[width=0.45\textwidth]{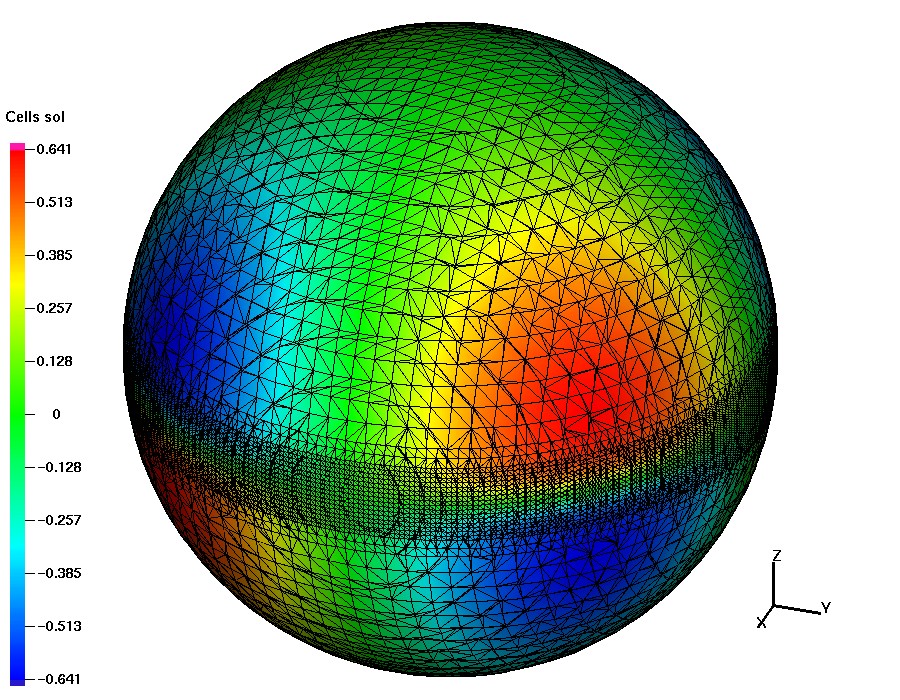}\qquad  \includegraphics[width=0.45\textwidth]{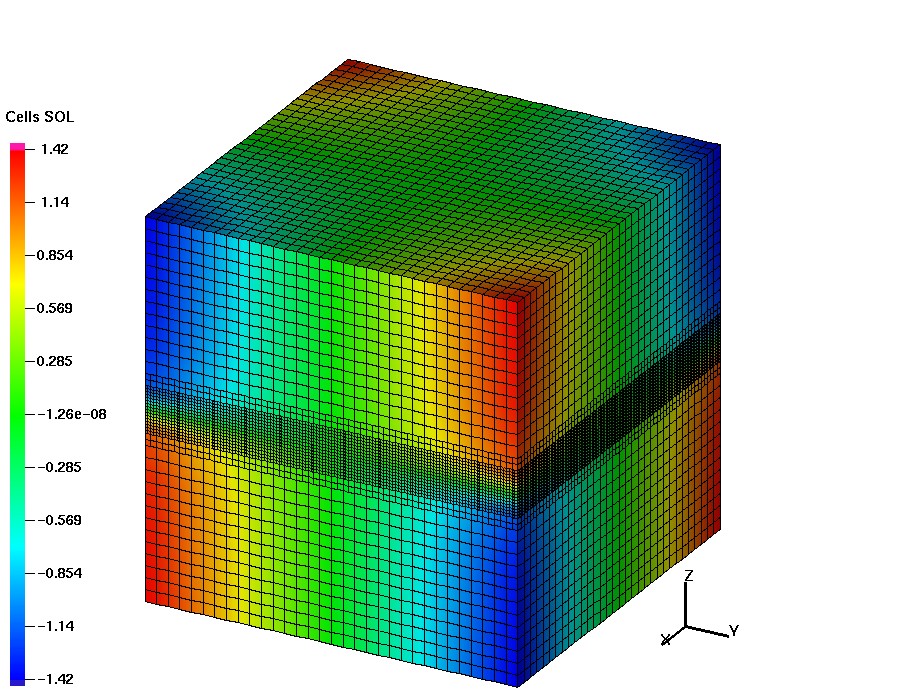}
\caption{\label{fig:sphere_adap}
Left: Induced surface mesh and the surface component of computed solution.
Right: The bulk mesh and the volume component of computed solution.}
\end{center}
\end{figure}

The material parameters are chosen as $D_1=D_2=D_\Gamma=\eps$ and $k_{1,a}=0.5$, $k_{2,a}=2$, $k_{1,d}=2$, $k_{2,d}=1$, $d=1$.
The source terms $f_i\in L^2(\Omega)$, $i=1,2$, and $f_\Gamma\in L^2(\Gamma)$   and  data on $\dO$ are taken such  that the exact solution of the  \textit{stationary} equations \eqref{diffeq1}--\eqref{diffeq2} is given by
\begin{equation*} %\label{exsol}
  %\begin{split}
  v(x,y,z) = {x z}\mathrm{arctan}\left(\frac{2z}{\sqrt{\varepsilon}}\right), \quad
  u_1(x,y,z) = 2 u_2(x,y,z), \quad
  u_2(x,y,z) = e^{1-x^2-y^2-z^2} v(x,y,z).
%\end{split}
\end{equation*}
We take $\eps=1$ (very smooth solution) and $\eps=0.01$ (solution has an internal layer along the midplane $z=0$).

\begin{table}
\begin{center}
\caption{Convergence of numerical solutions in the experiment with a spherical $\Gamma$ and locally refined mesh.
\label{tab:sphere_adap}  }\smallskip
\small
\begin{tabular}{r|llllll|llllll}\hline
&\multicolumn{6}{|c|}{$\varepsilon=1$}&\multicolumn{6}{|c}{$\varepsilon=1e-2$}\\
\#d.o.f. & $L^2$-norm & rate & $H^1$-norm& rate & $L^\infty$-norm& rate& $L^2$-norm & rate & $H^1$-norm& rate & $L^\infty$-norm& rate \\ \hline\\[-2ex]
{$\Omega$}&&&&&&&&&&&&\\
5840  &   3.993e-03&       &  8.322e-02&        &  1.706e-02&      &   5.666e-02&       &  3.987e-01&        &  1.945e-01&     \\
43552 &   6.706e-04&2.57   &  2.828e-02&1.56    &  4.926e-03& 1.79 &   2.609e-02&1.12   &  2.440e-01&0.71    &  8.403e-02&1.21 \\
318696&   2.200e-04&1.61   &  1.038e-02&1.45    &  2.158e-02&-2.13 &   1.353e-02&0.95   &  1.708e-01&0.52    &  5.003e-02&0.75 \\  \hline\\[-2ex]
{$\Gamma$} &&&&&&&&&&&&\\
1500 &       1.916e-03&    &   3.609e-02&     & 6.850e-03&    &    8.353e-03&    &  4.026e-01&     & 4.538e-02&     \\
6740 &       5.106e-04&1.91&   1.710e-02&1.08 & 1.919e-03&1.84&    1.854e-03&2.17&  1.619e-01&1.31 & 1.335e-02&1.76 \\
25988&       1.400e-04&1.87&   8.924e-03&0.94 & 5.624e-04&1.77&    3.848e-04&2.26&  6.532e-02&1.31 & 3.694e-03&1.85 \\  \hline
\end{tabular}
\end{center}
\end{table}

We build a sequence of locally refined meshes as illustrated in~Figure~\ref{fig:sphere_adap}.
For $\eps=0.01$ \rev{the meshes are fitted to the layer} and intend  to capture the sharp variation of the solution.
We computed numerical solutions on a sequence of 3 meshes, the second mesh is illustrated in ~Figure~\ref{fig:sphere_adap}. Each mesh has two levels of refinement in the region $|z|<\frac18$.
The convergence of the method is reported in Table~\ref{tab:sphere_adap}. The optimal order of convergence is attended for the surface component of the solution, but the FV method in the bulk domain shows lower order convergence for the convection dominated case. We conclude that more studies are required to improve the performance of the FV method on such type of meshes.

\section{Conclusions} \label{s_concl} The paper proposed a hybrid finite volume -- finite element method for the coupled bulk--surface systems of PDEs. The distinct feature of the method is that the same background mesh is used to solve equations in the bulk and on the surfaces, and that there is no need to fit this mesh to the embedded surfaces. This makes the approach particularly attractive to treat problems with complicated embedded structures of lower dimension like those occurring in the simulations of flow and transport in fractured porous media. We consider the particular monotone non-linear FV method with compact stencil, but we believe that the approach can be carried over and used with other FV methods on polyhedral meshes (e.g. some of those reviewed in \cite{Droniou:14}) with possibly better performance in terms of convergence rates. In this paper we treated only diffusion and transport of a contaminant assuming that Darcy velocity is given. Extending the method to computing flows in fractured porous media is in our future plans together with the design of better algebraic solvers, doing research on adaptivity, and adding to the method a fracture prorogation model.

}

\subsection*{Acknowledgments}
The work of the first author (the numerical implementation and the numerical experiments) has been supported  by the Russian Scientific Foundation Grant 17-71-10173, the work of the second author has been supported by the NSF grant 1717516, the work of the third author has been supported by the RFBR grant 17-01-00886.

\bibliographystyle{abbrv}
\bibliography{ChernOlshan}
\end{document}